\documentclass[aps,pra,twocolumn,superscriptaddress,floatfix]{revtex4-2}

\usepackage{graphicx}
\usepackage{subfigure}
\usepackage{amsmath}
\usepackage{mathrsfs}
\usepackage{amsthm}
\usepackage{amssymb}
\usepackage{hyperref}
\usepackage{xcolor}
\usepackage[OT4]{fontenc}
\usepackage[utf8]{inputenc}

\usepackage{ulem}
\usepackage{dcolumn}
\usepackage{bm}

\hypersetup{colorlinks=true, linkcolor=black, citecolor=black, urlcolor=blue}
\allowdisplaybreaks

\begin{document}

\newcommand{\x}{{\bf r}}
\newcommand{\K}{{\bf k}}
\newcommand{\q}{{\bf q}}
\newcommand{\dk}{  \Delta {\bf k}}
\newcommand{\DK}{\Delta {\bf K}}
\newcommand{\KK}{{\bf K}}
\newcommand{\X}{{\bf R}}

\newcommand{\B}[1]{\mathbf{#1}} 
\newcommand{\f}[1]{\textrm{#1}} 

\newcommand{\half}{{\frac{1}{2}}}

\newcommand{\vv}{{\bf v}}
\newcommand{\p}{{\bf p}}

\newcommand{\dx}{\Delta {\bf r}}

\author{Maciej Pylak}
\affiliation{National Centre for Nuclear Research, ul. Pasteura 7, PL-02-093 Warsaw, Poland}
\email{maciej.pylak@ncbj.gov.pl}

\author{Mariusz Gajda}
\affiliation{Institute of Physics, Polish Academy of Sciences, Aleja Lotnik\'ow 32/46, PL-02-668 Warsaw, Poland}

\author{Pawe{\l} Zin}
\affiliation{National Centre for Nuclear Research, ul. Pasteura 7, PL-02-093 Warsaw, Poland}

\title{Dipolar Droplets at 3D-1D Crossover}

  \begin{abstract}
We investigate beyond-mean-field corrections to the energy of an elongated homogeneous Bose gas strongly confined in two directions, with dipoles aligned along the long axis of the system. When the dipolar interaction reaches its critical strength, the mean-field approach predicts instability. However, similar to the free-space case, beyond-mean-field effects significantly alter the ground state of the system, leading to the formation of a self-bound atomic cloud known as a quantum droplet. Our analysis demonstrates that the beyond-mean-field contribution to the energy in the quasi-1D region, in addition to the confinement induced shift of the mean field energy, is proportional to the third power of the density $\sim n^3$. Therefore, it can be interpreted as an effective three-body repulsion that stabilizes the gas, preventing collapse and leading to a finite-density solution. We also show  that the same effect plays a crucial role in the binding of strongly elongated dipolar droplets under harmonic confinement.
  
\end{abstract}

\maketitle

\section{Introduction}
Ultracold gases in the regime of Bose-Einstein condensation have been the focus of interest for researchers for many years. This new state of matter allows for the observation of new exotic phenomena. Particularly interesting in this context are dipolar gases. In addition to short-range and isotropic contact interactions, they also experience a long-range anisotropic potential, giving rise to many new effects. A review of the main advances in this field can be found in \cite{baranov2012,Lahaye:2009}.

Such systems have been studied for magnetic atoms of chromium \cite{Goral2000}, polar molecules \cite{bohn2017cold} and Rydberg atoms \cite{saffman2010quantum}. Strongly dipolar Bose-Einstein condensates (BEC) have been experimentally realized in dysprosium \cite{lu2011strongly} and erbium atoms \cite{aikawa2012bose}. It was also shown that the system geometry plays a crucial role in the stability of the cloud. Advances in experimental techniques triggered further research, resulting in the discovery of self-bound systems of dipolar atoms called quantum droplets \cite{Kadau2016,Barbut2016,Chomaz2016,Schmitt2016,Wenzel2017}. This achievement was made possible by fine-tuning the scattering length, accomplished through Feshbach resonances. As a result, the contact repulsion competes with dipolar attraction, and these two terms cancel each other out, causing the mean-field contribution to the total energy to vanish. In this scenario, it becomes necessary to include beyond mean-field terms, which are typically omitted as small corrections, as first noticed by D.~Petrov, \cite{Petrov2015} in a somewhat different system, i.e. in a two component mixture of Bose-Bose atoms. Without these corrections, the system would be unstable, tending to collapse. The mentioned experiments encourage theoretical investigations into these beyond mean-field corrections. 

The pioneering papers \cite{Lee1957,Beliaev1958,Schick1971,hugenholtz1959ground} suggest considering the positive contribution to the chemical potential, believed to stabilize the system and allowing for droplets formation. For a homogeneous Bose gas with a repulsive contact potential, this repulsive term is known as the Lee-Huang-Yang (LHY) term \cite{Lee1957}. A similar correction has been found  in \cite{Uwe2006} for dilute Bose–Einstein condensates with two-particle interaction potentials and for a homogeneous dipolar gas \cite{Pelster2012}. It is not surprising that the density dependence of the LHY term in dipolar systems is the same as for the contact interactions; however, its magnitude is enhanced by the strength of the dipolar interaction. 
The majority of theoretical investigations in droplet's physics were restricted to finding of a solution to the extended Gross-Pitaevskii (eGP) equation with an additional term accounting for the LHY correction \cite{Wachtler2016,bisset2016ground} for homogeneous 3-dimensional dipolar gas 
\cite{Pelster2012}. It must be stressed that experimental systems are not homogeneous, therefore  the local density approximation (LDA) is a crucial factor  in verifying this technique. Until now,  predictions of the extended GP equation have been quite effective in understanding the experimental results. They are also consistent with Monte Carlo calculations \cite{Macia2016droplets, Cinti2017, Bottcher2019a}, however, some departures from the mean field predictions for squeezed droplets were studied recently \cite{Sanuy2024}.

Confinement of the gas in an external potential leads to an interesting behavior \cite{Popov,Mora2009,Zin_dipole}, including crucial changes in the excitation spectrum and the development of roton modes \cite{Santos2003,Fischer2006,Boudjemaa2013,Chomaz2018,Petter2018,Kora2019}.
Therefore dimensionality of the system and its geometry strongly affects its  physical properties by introducing substantial modifications of the beyond mean field contribution to the energy,  in the region of a weak collapse.

The LHY energy for ultradilute two-component liquids with zero-range interactions in  1D and 2D geometric arrangements is determined in \cite{Petrov2016}. In 3D systems the effects of confinement on the LHY energy have been studied in both two-component systems with zero-range interactions \cite{zin2018droplets,buechler2018crossover,Zin2022a}, as well as in systems with dipole-dipole interactions \cite{Edler2017, Zin_dipole}. In the case of a uniform system with zero-range interactions, the LHY energy at the 3D-2D and 3D-1D crossovers is found in \cite{zin2018droplets,buechler2018crossover}. The same systems, but under harmonic confinement, are also studied: the 3D-1D crossover   is considered in \cite{buechler2018crossover} where the local density approximation is assumed, while the 3D-2D case is investigated in \cite{Zin2022a}. In the latter case, no local density approximation in the confined direction is assumed, and a modified gapless Hartree-Fock-Bogoliubov method is  formulated to treat such systems \cite{Zin2022b}. Single-component systems with long-range dipole-dipole interactions have been studied in a quasi-2D geometry, assuming a homogeneous density profile \cite{Zin_dipole}.

An inhomogeneous system at the 3D-1D transition, squeezed by a harmonic potential, is considered in \cite{Edler2017}. Here, we focus on a similar scenario, but instead of harmonic confinement, we assume a dipolar Bose gas that is homogeneous in the transverse plane and confined within a square box of size $L \times L$. The system we study has a pencil-like shape, making it effectively one-dimensional. Our goal is to determine the Lee-Huang-Yang (LHY) energy correction in this setup.
While more realistic confining potentials generally lead to nonuniform density profiles, the characteristic dependence of the LHY energy term on the density is, to a large extent, universal. As such, homogeneous systems serve as a crucial tool for understanding higher-order corrections to the energy.


The manuscript is organized as follows. In Section \ref{sec:system} we introduce  the system and discuss basic assumptions as well as  essential energy and length scales involved. Then we present the expression for the LHY energy, and show our approach to find the interaction potential reduced  to the confined geometry.  In this section we also identify the region of the system stability.  In Section \ref{sec:LHY} we compare different limiting physical situations.  We present analytical, to a large extend,  expression in the quasi-1D limit, and compare the result to the one obtained for the  1D geometry.   We also compare our findings to the results obtained in the situation when confinement is provided by a harmonic potential.  The stabilizing effect of the LHY energy and its role in formation of a dipolar strongly elongated droplet is discussed  in section \ref{sec:dipolar_droplet}.   Conclusions are presented in section \ref{sec:conclusions}.

\section{Lee-Huang-Yang  energy at 3D-1D crossover}
\label{sec:system}

\subsection{Mean field energy functional}
Although, strictly low-dimensional systems are not reachable experimentally, typically, a very steep external potential is applied to confine  atoms in one or two dimensions.  Resulting systems are three-dimensional, but   excitation energies in the "tight" direction(s) are very large, significantly exceeding  excitation energies in the unconfined direction(s). At sufficiently low temperatures, such squeezed systems, having  pencil- or a disc-like shapes, are in the ground state of the confinement.  Their kinematics, characterized by a {\it continuous} excitation spectrum in the "free" direction(s) and  {\it discrete} excitation energies in the confined space, is effectively low dimensional. Such systems are at  dimensional crossover. It is only when a contribution to the energy from discrete excitations  becomes negligible, when the system reaches  the quasi-1D (quasi-2D) regime.


If we assume that confinement is provided by a potential that is infinite at the square box walls in the $x$-$y$ plane, the wave function of the system is zero on the box boundaries, leading to the problem of a nonuniform density profile. In such cases, numerical calculations are typically required. However, we do not wish to proceed in this manner. Our goal is to find analytical expressions. Therefore, to reconcile confinement with a uniform density profile, we explicitly impose periodic boundary conditions in the $x$-$y$ plane. As a result, our system takes on the topology of a two-dimensional torus in the $x$-$y$ plane. Thus, whenever we refer to the confined system or box-like confinement, we mean a uniform system with periodic boundary conditions in the $x$-$y$ plane, with a period of $L$ in both directions.

Studying such `toy models' is mathematically well-defined and commonly used in similar contexts \cite{Petter2018, zin2018droplets,Zin_dipole}. The availability of analytic expressions for the Bogoliubov excitations and their eigenmodes provides a significant simplification. The results obtained for a uniform system exhibit general features that are independent of the confining potential. We will leverage this property in subsequent sections.

A density of the system, uniform in the $x$-$y$ plane, might vary in the $z$ direction,   i.e. it has the form of:
\begin{equation}
\label{dens3D}
    n =\frac{1}{L^2}n_{z}(z)= |\psi(z)|^2,
\end{equation}
where $n_{z}(z)$ is the one-dimensional density related to the wavefunction $\psi(z)$, $n_z(z)=L^2 |\psi(z)|^2$.

Assuming that the system is in the Bose-Einstein condensate  phase, its mean field energy functional depends on the order parameter $\psi(z)$: 
\begin{widetext}
\begin{eqnarray}
      E_{mf}[\psi]&=&\int dz \int_{L\times L}  d\x_\perp  \left( \frac{\hbar^2}{2m} |\partial_z \psi|^2+\frac{1}{2} \int dz' \int_{L\times L}  d\x_\perp'   v(\x-\x') |\psi(z)|^2 |\psi(z')|^2\right),
  \label{eq:energia_1d}
\end{eqnarray}
\end{widetext}
where $v(\x)$ is the two-particle interaction potential that obeys periodic boundary conditions in the $x$-$y$ plane, to assure consistency with our assumptions about periodic boundary conditions imposed on the system. Therefore the potential should be defined on a 2D torus as $x$ and $y$ dependence is concern.  The periodic potential $v(\x)$ can be conveniently introduced via a discrete Fourier transform $v(\K)$:
\begin{equation}
\label{eq:periodic}
    v(\x)= \frac{g}{2\pi L^2} \sum_{k_x,k_y} \int \mbox{d}k_z e^{ i k_x x+ i k_y y}    e^{i k_z z} v(\K),
\end{equation}
where $k_x$ and $k_y$ are components of the wavevectors $\K$ which take discrete values: $k_{x,y}=(2\pi/L) q_{x,y}$, and $q_x,q_y \in \mathbb{Z}$. 

If the potential is short-range, the boundary conditions imposed on the system do not affect its form. However, the problem arises when the interactions are long-range. In the following, we specify the form of the potential $v(\x)$ on the torus in the case where the 3D interactions involve long-range dipole-dipole interactions.

\subsection{Interactions accounting for periodic boundary conditions}
 
 In the case of the system we study here, the interaction potential in the free 3D space,  $v_{3d}(\x)$,  consists of two main parts: i) the dipole-dipole anisotropic interaction, and ii) a short range isotropic repulsion which dominates over the dipolar part at small distances. At  low energies  and  densities the adequate form of the interaction potential is: \cite{Yi2000,Oldziejewski2016}:
\begin{equation}
\label{eq:v}
v_{3d}(\x)=g\left[\delta(\mathbf{r})+\frac{3\varepsilon_{dd}}{4\pi}\frac{\left(1-3 ({\bf e}_\x \cdot {\bf e})^2 \right)}{r^3}\right].
\end{equation} 
The short range interaction strength $g=\frac{4\pi\hbar^2a}{m}$, is determined by the scattering length, $a$, and atomic mass, $m$, while  $\varepsilon_{dd}=\frac{a_{dd}}{a},$ determines the strength of dipole-dipole forces, and $a_{dd}=\mu_0|\mu|^2m/(12\pi\hbar^2)$, is  a characteristic length of dipolar potential,  where $\mu_0$ is vacuum permittivity, $\mu$ is dipolar moment of an atom and $m$ its mass. In the above equation ${\bf e}_\x = \x/|\x|$, is a versor in the $\x$ direction and ${\bf e}$, is the unit vector in the direction of the dipole moment, i.e. the z-direction. 
The Fourier transform of the 3D interaction potential:
\begin{equation}
\label{eq:fourier}
  v_{3d}(\K) = \frac{1}{g} \int d \x \, e^{-i\K \x} v_{3d}(\x) =1+\varepsilon_{dd}\left(3\frac{k_z^2}{k^2}-1\right),  
\end{equation}
involves all real values of wavector $\K=(k_x,k_y,k_z)$.  The factor $1/g$ is introduced for the future convenience.   
The simplest choice of $v(\K)$ leading to a periodic $v(\x)$ is to chose   $v(\K)= v_{3d}(\K)$ in Eq.~(\ref{eq:periodic}). The integration in Eq.~(\ref{eq:fourier}) is carried over entire $x$-$y$ plane, thus the Fourier transform of the potential involves contributions from infinitely many periodic copies of the unit cell. This choice corresponds to the infinite periodic system of "square tubes". 

Our system is a single tube with periodic boundary conditions in the vertical plane. 
The interaction potential should account for contributions from the unit cell of size $L \times L$ only:
\begin{equation}
\label{gvk}
   v(\K) = \frac{1}{g}\int_{L\times L}{d\x_\perp \int{dz\, e^{-i \K \x} v_{3d}(\x)}}.
\end{equation}
The real-space potential $v(\x)$ is given by the inverse Fourier transform, Eq.~(\ref{eq:periodic}). Clearly, $v(\x)$ differs from $v_{3d}(\x)$. It is defined on a 2D torus in the vertical plane and extends along the entire longitudinal $z$-axis. The range of interactions in the $x$-$y$ plane is limited by the spatial size of the confinement.

Using the dipole-dipole potential "tailored" to the toroidal geometry introduces a spurious contribution to the interaction energy of the system, as the interactions between two atoms on a torus differ from those in free 3D space. However, in Appendix~\ref{App:D}, we show that the difference between the interaction energies for the 3D potential, $v_{3d}(\x)$, and the periodic one, $v(\x)$, is very small. Numerical estimates based on the equations in Appendix~\ref{App:D} indicate that this difference is below 1\%, provided that the density profile in the $z$-direction varies over distances much larger than the healing length, $l_h$.

The use of the potential $v(\x)$ is necessary to ensure the model is consistent and well-defined, as it guarantees the correct physical properties of the system. As shown in \cite{Edler2017}, the confined system becomes unstable within the mean-field formalism if the strength of the dipole-dipole interaction exceeds one, $\varepsilon_{dd} > 1$. Although \cite{Edler2017} assumes harmonic confinement in the $z$-direction, the collapse instability should be attributed to inter-particle interactions rather than the specific shape of the wave function in the $z$-direction imposed by the confinement. Therefore, the value $\varepsilon_{dd} = 1$ defines the mean-field instability point, independent of the confinement.

In the following sections, we show that indeed, for the model studied here, the Bogoliubov excitation spectrum of the system with interactions given by $v(\x)$ exhibits the same kind of instability, which is triggered when $\varepsilon_{dd} = 1$. 



\subsection{Physical assumptions}
The characteristic length scale associated with the variation of density — the healing length $l_h$ — is determined by interactions. As studied in detail in \cite{Abad2020}, the healing length of an oblate condensate with strong dipole-dipole interactions, comparable to short-range repulsion, increases, compared to that of a standard repulsive condensate, by a factor $\beta$, depending on the scattering length and system geometry: $l_h \approx \beta \times 1/\sqrt{8 \pi a n}$. The healing length reaches its maximum at the transition point from the superfluid to droplet regime and, for typical parameter values \cite{Abad2020}, can increase several times compared to its value in the superfluid phase.

We assume that the healing length, particularly in the quasi-1D region, is much larger than the transverse  extension of the system, $l_h \gg L$, but much smaller than its longitudinal size, $l_h \ll L_z$. The latter condition allows the use of the local density approximation, replacing the constant density with the local density $n_{z}(z)$, which varies slowly with $z$.

Consistently with the above assumptions, transverse momentum is quantized and takes discrete values $\K=\frac{2\pi}{L} \q$, where $q_x,q_y \in \mathbb{Z}$, and one-particle  excitation spectrum in the $x$-$y$ plane equals to $\varepsilon_k=\frac{\hbar^2k^2}{2m}$.   The ratio $\xi$, of the characteristic mean field energy $g n$ ($n$ is atom density), to the elementary excitation energy in the confined plane, $\epsilon_0 = \frac{\hbar^2}{2m} (\frac{2\pi}{L})^2$, is the fundamental  parameter of our study:  
\begin{eqnarray}
    \xi = \frac{gn}{\epsilon_0} = \frac{2a}{\pi} (n L^2) =\frac{2}{\pi }a n_{1d}.
\end{eqnarray}
In the above equation, if the system is not uniform, the density $n$ should be chosen as the typical density, for example, at its center, and $n_{1d} = n L^2$ is the corresponding 1D density, $n_{1d}=n_z(0)$. The parameter $\xi$ plays a crucial role in determining the effective dimensionality of the system: $\xi \gg 1$ signifies a 3D geometry, $\xi < 1$ corresponds to an elongated system, while $\xi \ll 1$ characterizes a quasi-1D system where longitudinal excitations dominate at low energies. We will refer to the system as being at a dimensional crossover when $\xi < 1$, and use the term quasi-1D if $\xi  \ll 1$. One has to remember however, that if the density becomes too small,   the 1D (quasi-1D) system  enters  strongly interacting Tonks-Girardeau phase. Because in our studies we implicitly assume  weakly interacting regime,  the  condition   $n_{1d} > 2\pi  (a/L^2)$ has to be fulfilled,  so the parameter $\xi$ is limited from below $\xi > 4 (a/L)^2$,  \cite{buechler2018crossover}.

There are two additional important length scales directly related to inter-particle interactions. The s-wave scattering length, $a$, determines the two-atom scattering cross section due to their short-range interactions. We assume that the linear size of the 2D confinement is much larger than the scattering length, i.e., $a \ll L$, i.e. collisions between atoms have a 3D character and  interactions are governed by 3D processes across the entire crossover range of $\xi$, including the quasi-1D region.

Another key length scale, the dipolar length $a_{dd}$, is introduced by dipole-dipole interactions, related to the dipolar scattering cross-section, $\propto a_{dd}^2$. Liquid droplets form when the mean-field energy becomes negligible, which occurs when dipole-dipole and short-range interactions contribute equally to the total energy and cancel each other out. This balance happens when the dipolar and scattering lengths are approximately equal, i.e., $a_{dd} \sim a \ll L$. 



\subsection{LHY energy density}
\label{sec:lhy}

When the mean-field energy of the system vanishes, the higher-order correction to the energy, namely the Lee-Huang-Yang (LHY) energy, becomes significant. In 3D space, the LHY energy density, $e_{LHY}(\xi)$, expressed in units of $\epsilon_0 / L^3$, can be derived following the reasoning presented in \cite{hugenholtz1959ground} (see also \cite{Petrov2015, Zin2022b}). It is given by:
\begin{equation}
  \label{eq:lhydef_3D}
-\frac{ 2 e_{LHY}^{3d}}{\xi^2}= 
  \int\,d\q  \frac{v_{3d}^2(\q)}{ \varepsilon_{3d}(\q) + q^2 + \xi v_{3d}(\q)} 
  -\int\,d\q \frac{ v_{3\mathrm{d}}^2(\q)}{2q^2},
\end{equation}
where $\varepsilon_{3d}(\q)$: 
\begin{equation}
\label{eq:bogol}
\varepsilon_{3d}(\q)=\sqrt{q^2[q^2+2\xi v_{3d}(\q)]},    
\end{equation}
are  Bogoliubov excitation energies in 3D. $v_{3d}(\q)$ is the Fourier transform of the interaction potential, Eq.~(\ref{eq:v}), and is given by Eq.~(\ref{eq:fourier}). Continuous wavector $\q=(q_x,q_y,q_z)$ is dimensionless. It is related to the wavevector $\K=(k_x,k_y,k_z)$ by the relation, $\K=\frac{2\pi}{L} \q$.

The first integral in right hand side of the Eq.~(\ref{eq:lhydef_3D}), represents cumulative   zero-point energy  of  Bogoliubov quasiparticles,  while the next one originates in the contribution to the interaction strength $g$, from the second order term in the Born expansion of the $T$-matrix.   Its presence  is important because it removes ultraviolet divergence of the first term (for details see \cite{Zin2022b}). 

If the system is confined, corresponding excitation spectrum is  discrete. Therefore, to account for the elongated geometry, the first integral in Eq.~(\ref{eq:lhydef_3D}) has to be replaced by a sum over discrete quantum numbers enumerating eigenstates in the confined subspace,  and integral over continues eigenvectors. The second term, which stems from atom-atom scattering, should remain unchanged, because the scattering in the whole  cross-over region is the 3D process, according to the assumption, $a \ll L$. Therefore, the LHY energy density at 3D-1D crossover region, $e_{LHY}^{1d}(\xi)$,  is given by:
\begin{equation}
  \label{eq:lhydef}
  -\frac{2 e_{LHY}^{1d}}{\xi^2}=\sum_{q_x,q_y=0}\int\,dq_z   \frac{v^2(\q)}{ \varepsilon(\q) + q^2 + 
  \xi v(\q)}-\int\!\! d\q \frac{ v_{3\mathrm{d}}^2(\q)}{2q^2},
\end{equation}
where in the first term  of  Eq.~(\ref{eq:lhydef}) we substitute the Fourier transform of  $v_{3d}(\q)$, by the Fourier transform of a modified interaction potential $v(\q)$, accounting for the confinement. This substitution  affects also the Bogoliubov excitation energies:
\begin{equation}
\label{eq:bogol_1d}
\varepsilon(\q)=\sqrt{q^2[q^2+2\xi v(\q)]},   
\end{equation}

The Fourier transform of the interaction potential $v(\mathbf{q})$ can be brought to the  following form:
\begin{widetext}
\begin{eqnarray}
\label{eq:vq}
\hspace*{-2cm}v({\bf q}) =  1 - \varepsilon_{dd}+3 \frac{\varepsilon_{dd}  q_z^2}{\pi}  \int{d q_x' \, \frac{\sin (\pi(q_x - q_x'))}{ q_x - q_x'}\cdot\frac{1 - (-1)^{q_y} \exp \left( - \pi \sqrt{{q_x'}^2 + q_z^2} \right) }{{q_x'}^2+q_y^2 + q_z^2}}.
\end{eqnarray}
\end{widetext}
Details of calculations are presented in  Appendix~\ref{app0}.
This expression allows to determine numerically, both the Bogoliubov excitation spectrum as well as  the LHY energy density at the crossover region. 

Analysis of the Bogoliubov excitation spectrum $\varepsilon(\mathbf{q})=\sqrt{q^2(q^2+2\xi v(\mathbf{q}))}$, leads to prediction of  instability of the system. The eigenenergies  become imaginary,  if   $v(\mathbf{q})<-\frac{q^2}{2\xi}$.  The integral in Eq. (\ref{eq:vq}) is equal to zero for $q_z=0$ and is positive otherwise. The most vulnerable  is the region of low momenta thus. We find that for $q_x, q_y=0$, and $q_z\to 0$,  the potential equals to:  $v(\mathbf{q})=1-\varepsilon_{dd}$, therefore the instability appears if:
\begin{equation}
\label{stability}
\varepsilon_{dd} \equiv \varepsilon_{dd}^{cr} = 1.    
\end{equation}
This is in fact the value given by the mean-field analysis as expected. 
Physical consequences of this fact are very important: at $\epsilon_{dd}^{cr}=1$, the interaction between atoms vanishes at $q=0$ and remains small in the neighborhood. Therefore the excitation energies at $\varepsilon_{dd}^{cr}$ are very small. 
This is in contrast to the case of two-component droplets with short-range interactions. In these systems, only the soft mode excitation energies are small, but not excitations of the hard ones.

\section{Limiting cases and comparison to other results}
\label{sec:LHY}
\subsection{Quasi-1D limit}

First, we discuss the regime where $\xi\ll 1$, which corresponds to the quasi-1D limit. The system is diluted enough for collisions to occur in three dimensions, but the population of excited states in the confined plane is low. As was shown in Appendix~\ref{app:eLHY1d}, the energy density of the LHY term, Eq.(\ref{eq:lhydef}) can be approximated as:
\begin{eqnarray}
  \label{eq:lhy1d}
  e_{LHY}^{1d}(\xi) \simeq  c_2 \xi^2 + c_3 \xi^3,
\end{eqnarray}
where:
\begin{eqnarray}\label{c2}
    c_2&=&\frac{1}{2}\left(\sum_{q_x,q_y=0}\int{dq_z\,\frac{v^2(\q)}{2q^2}}-\int{d\q\,\frac{v_{3d}^2(\q)}{2q^2}}\right) \nonumber \\
    &&\simeq-0.479, \\
    c_3&=&\frac{1}{2} \sum_{q_x,q_y=0} \int \mbox{d} q_z \, \frac{v^3({\bf q})}{2q^4} \simeq 84.5301.
    \label{c3}
\end{eqnarray}
In obtaining coefficients $c_2, c_3$ we approximated the value of $\varepsilon_{dd}$  by its value at the critical point, $\varepsilon_{dd}=1$, which seems reasonable. As demonstrated in \cite{Ota2020,Hu2020}, the LHY term changes very weakly when $\varepsilon_{dd}$ crosses unity and enters the critical regime.  

The first term $\propto \xi^2$, Eq.~(\ref{c2}), arises from the external confinement, reflecting its influence on two-body scattering. A similar contribution has been identified in previous studies \cite{zin2018droplets, buechler2018crossover}, where dimensional crossover in the context of quasi-2D dipolar gases was examined \cite{Zin_dipole}. However, in the present case, this term is negative, $c_2<0$, unlike in the 2D geometry, thereby failing to ensure the stabilization of the droplet. This term, quadratic in the atom density, has the same character as the contribution from the contact interactions. Therefore it   can be absorbed into the scattering length, shifting  the critical value of the dipole-dipole interactions strength, by a relatively small value  $\frac{4 a}{\pi L} c_2$, i.e. $\varepsilon_{dd}^{cr}=1 \to \varepsilon_{dd}^{cr}=1+\frac{4 a}{\pi L} c_2$. This effect is known as the confinement-induced resonance, \cite{Olshanii98,Olshanii03}.


The second term, Eq.~(\ref{c3}), proportional to $\propto \xi^3$,   stems from three-body processes inherent in quantum fluctuations.  Since this term indicates repulsion between particles, $c_3>0$, it provides a mechanism for stabilization and formation of quantum droplets in quasi-1D geometry. Although being of the higher order than  the term $\propto c_2$, it is the most important, nontrivial contribution to the LHY energy. 

\begin{figure}[htb]
    \centering
 \includegraphics[width=0.95\linewidth]{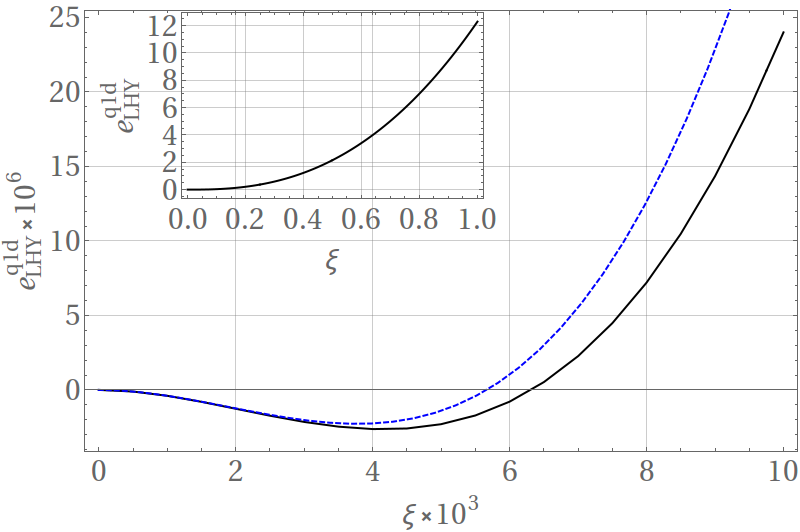}
    \caption{The LHY energy $e_{LHY}^{1d}$ as a function of $\xi$ at the critical point, $\varepsilon_{dd}=1$. Black line indicates obtained  numerically LHY energy according to Eq. (\ref{eq:lhydef}) and blue dashed line shows approximate result given by Eq. (\ref{eq:lhy1d}).  Inset: high density region of  the LHY energies exhibiting 3D limit behavior given by Eq. (\ref{eq:lhy3d}).}
    \label{fig:eLHY}
\end{figure}

As illustrated in Fig. (\ref{fig:eLHY}),  approximation of the Eq. (\ref{eq:lhydef}) by the semi-analytic expression, Eq. (\ref{eq:lhy1d}), gives correct behavior of the LHY energy if $\xi\lesssim0.003$. Outside this region, numerical integration is necessary, unless the parameter $\xi$ becomes of the order of one. 

For $\xi> 1$ not only the states in the `free' \textit{z} direction, but also the states in the confined \textit{x,y} directions, are significantly occupied. In consequence the system behaves as three-dimensional,  therefore, it is not surprising, that Eq~(\ref{eq:lhy1d}) recovers the  well known 3D result:
\begin{equation}
    \label{eq:lhy3d}
    e_{LHY}^{1d}(\xi)\stackrel{\xi\to\infty}{\longrightarrow}e_{LHY}^{3{d}}= \frac{8\pi \sqrt{6}}{5}\xi^{5/2}
\end{equation}

\subsection{1D limit}
\label{sec:1D_limit}
Our results  obtained for a very elongated sample, provide a bridge between the 3D and quasi-1D systems. 
It is shown in \cite{zin2018droplets,buechler2018crossover} that 
the leading term contributing to the LHY energy 
in the case of two-component systems with contact interaction,  both in the quasi-1D as well as in 1D geometry,  are identical.  

In the single component dipolar systems, the first essential term of the LHY energy, in 1D and quasi-1D geometry, is proportional  to $c_3 \xi^3$ or $c^{1d}_{3}\xi^3$ respectively, where coefficients $c_3$ or $c_3^{1d}$ are:
\begin{eqnarray}
 c_3 &=& \frac{1}{2} \sum_{q_x,q_y=0} \int \mbox{d} q_z \, \frac{v^3({\bf q})}{2q^4} \simeq 84.5301,
\\
c^{1d}_{3} &=& \frac{1}{2} \int \mbox{d} q_z \, \frac{v^3(q_z)}{2q_z^4} \simeq 76.8901.
\end{eqnarray}
These  terms, are evidently different: summation over perpendicular momenta, $q_x, q_y$, is  missing in 1D.  In the quasi-1D limit, we still have the impact of excited states with $q_x , q_y\neq 0 $ and this contribution to $c_3$ is not negligible. Numerical calculations indicate that  modes excited in the perpendicular plane contribute about $9\%$ of the total value.

Therefore quasi-1D and 1D limits are different for dipolar droplets. The reason  is that the Fourier transform of the  potential,  Eq.~(\ref{eq:vq}),  vanishes in the limit of low momenta, $q_z \rightarrow 0$ for $q_x=q_y=0$, and, what is crucial, all  excitation energies at low momenta, responsible the LHY term, are small. 


The excitation spectrum of two-component droplets with repulsive intra-component and attractive inter-component contact interactions is different. The two components can oscillate together (in-phase), which corresponds to the soft modes characterized by low excitation energies, or they can oscillate against  each other (out-of-phase), which requires much more energy (hard modes). As the Fourier transform of the interactions responsible for the soft modes vanishes  when the droplets are formed, the analogous transform for the interactions governing the hard modes remains constant and equals the effective intra-component repulsion strength. This is why the hard modes dominate the LHY energy in such systems. The contribution from the soft modes is practically negligible and, in general, differs between quasi-1D and 1D systems.

Evidently in dipolar droplets hard modes are missing, only soft modes contribute to the LHY energy, and their contribution, similarly to the case of two-component droplets, differs between 1D and quasi-1D geometries, $c_3 \neq c_3^{1d}$.


\subsection{Harmonically trapped dipolar gas}
\label{subsec:harmonic}
It is interesting to compare our results to those obtained for similar systems, \cite{Petrov2016, Edler2017}. In the case of 1D sample with zero-range interactions, the LHY energy density is proportional to $e^{1D}_{LHY} \sim -\xi^{3/2}$, and it contributes to the chemical potential by an amount   $\Delta \mu_{LHY} \sim \frac{d}{d\xi} e^{1D}_{LHY}(\xi)  \sim -\xi^{1/2}$.

Only numerical calculations  are possible for quasi-1D system with  dipole-dipole interactions  confined by a harmonic potential.   It is concluded in \cite{Edler2017}, that the fit to numerical results, in the limit of $\xi \to 0$,  is consistent with  a quadratic contribution to the energy $e^{1D}_{LHY} \sim -\xi^2$, while for  large values of $\xi > 1$,  the typical LHY energy   in 3D geometry is recovered, $e^{1D}_{LHY} \sim \xi^{5/2}$. 
Noting that the quadratic term can be added to the mean-field energy, no significant contribution from confinement was observed in \cite{Edler2017}.  

Our results, obtained for the uniform gas, allow for more detailed studies due to the largely analytic nature of the findings. They reveal that, in addition to the terms identified in \cite{Edler2017}, there is also an important contribution proportional to $\xi^3$. This term should also appear in the expression for the LHY energy of a harmonically confined system, as the LHY energy arises from specific types of physical processes in reduced dimensions that are independent of the particular shape of the external potential providing the confinement. 

\begin{figure}[htb]
    \centering
 \includegraphics[width=0.95\linewidth]{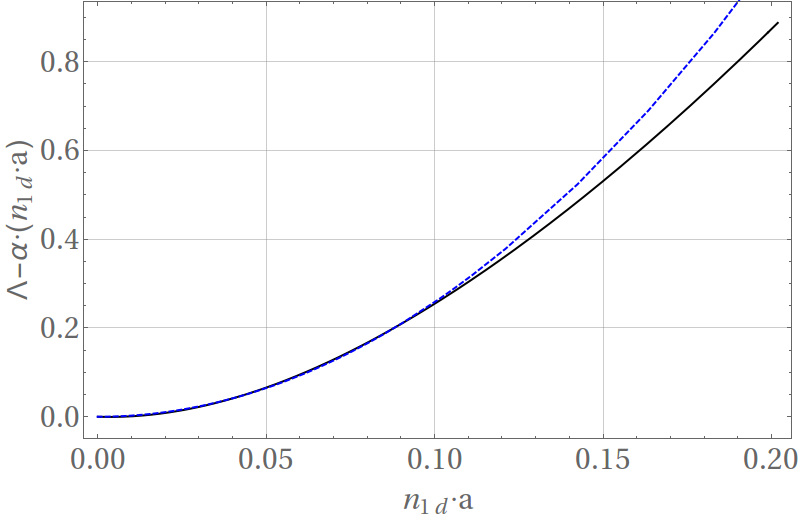}
    \caption{The LHY chemical potential $\Lambda=\frac{\Delta\mu_{LHY}}{\hbar\omega_\perp}\cdot\frac{L}{a}$ with subtracted linear term as a function of $n_{1d}a$. Black line indicates results presented in \cite{Edler2017}, blue dashed line shows contribution from quadratic term: $a_2\cdot (n_{1d}a)^2$, $a_2\simeq20.19$}
    \label{fig:santos}
\end{figure}

To show that this term is also present in results of \cite{Edler2017}, we analyze carefully  the data presented there. In Fig.~(\ref{fig:santos}) we plot the scaled chemical potential,  $\Lambda=\frac{\Delta \mu_{LHY}}{\hbar \omega_\perp}\frac{L}{a}$,  (black line),  as a function of $x=n_{1D}a \equiv \xi \pi/2 $, after subtracting the linear term first,  $\Lambda (x)-\alpha x$, originating in the quadratic contribution to the energy. The coefficient $\alpha$ is obtained by fitting to the results  of \cite{Edler2017}. The blue dashed line shows the term  $a_2 x^2$, coming form $\xi^3$ term in the LHY energy density,  where $a_2$ is fitted to the function $\Lambda(x)-\alpha x$. 

The agreement of our fits indicates that the third-order contribution to the LHY energy, $\sim \xi^3$, at small $\xi$, is also present in the numerical results shown in \cite{Edler2017}, but was not noticed there. The reason is that this term is overshadowed by the dominant quadratic (trivial) contribution to the LHY energy at small $\xi$. Our studies for a homogeneous system allow us to discover this missing contribution, which, although relatively small, determines the droplet density in quasi-1D geometry, what is discussed in the next section. 

\section{Dipolar quasi-1D droplet}
\label{sec:dipolar_droplet}



In this section, we analyze the possibility of forming a self-bound quantum droplet in an elongated geometry, and delve into their stability conditions and saturation density, particularly in the regime of large atom numbers.

The system is homogeneous in the $L \times L$ square in the $x$-$y$ plane; however, its density varies with $z$. The wavefunction can be approximated as $\psi = \frac{1}{L} \psi_z(z)$. Using this ansatz in Eq.~(\ref{eq:energia_1d}), and accounting for the LHY term, we obtain the effective 1D energy density functional, which depends on the longitudinal wavefunction $\psi_z$:
\begin{widetext}
\begin{eqnarray}
  E[\psi_z(\x)]=&&\frac{\hbar^2}{2m}\int dz\,|\partial_z \psi_z|^2 +\! \frac{1}{2}\int \!\!dz dz'\,   v_{1d}(z-z')
|\psi_z(z)|^2 |\psi_z(z)|^2 +\int d z \, \frac{\epsilon_0}{L} e_{LHY}^{1d}\!\left(|\psi_z(z)|^2 \right).
 \label{dlugie}
\end{eqnarray}
\end{widetext}
where we explicitly stress the dependence of the  $e_{LHY}^{1d}(|\psi_z(z)|^2)$ on the profile of the droplet in the longitudinal direction. 
The potential $v_{1d}$ present in the above equation is defined as
\begin{equation}\label{v1d:def}
    v_{1d}(z-z')=\frac{1}{L^4}\int_{L\times L}{v(\x-\x')\,d\x_\perp\,d\x_\perp'}.
\end{equation}


The energy given by Eq. (\ref{dlugie}) comprises three terms. When considering large droplets, the first term, representing the kinetic energy, can be disregarded, as only the edges, which have a length much smaller than the flat top, contribute significantly.  The next term is the dipole-dipole interaction energy, related to the effective potential, $v_{1d}(z-z')$. As it is shown in Appendix \ref{App:D}, it  can be approximated by the short-range one. Its characteristic range is small, being of the order of $L$, which is much less than a typical length scale of density variation. It turns out, that effective dipole-dipole interaction in Eq.~(\ref{dlugie})  can be substituted, to a good approximation, by the contact interaction: 
\begin{equation}\label{przyblizenie}
  v_{1d}\approx\frac{1-\varepsilon_{dd}}{L^2}g\delta(z-z').  
\end{equation}
Validity of this approximation was discussed  in details in quasi-2D geometry \cite{Zin_dipole}.

The third term in Eq. (\ref{dlugie}), the beyond mean field correction, involves the Lee-Huang-Yang (LHY) energy density, denoted as  $e_{LHY}^{1d}$. Eq. (\ref{eq:lhy1d}) is derived for the critical value of the parameter, $\varepsilon_{dd} = 1$. This is because the standard Bogoliubov approach fails in the unstable region, $\varepsilon_{dd} > 1$, due to the presence of imaginary components in the excitation energies.Therefore, in most studies, the  energy density $e_{LHY}^{1d}$, in the critical region is approximated by its value at the critical point. Consequently, we employ the form of the LHY correction given by the expansion into a series in the parameter $\xi$, for $\varepsilon_{dd}=1$, Eq.~(\ref{eq:lhy1d}), and,  according to the idea of the local density approximation,  $\xi$ is substituted by  $\xi = \frac{2}{\pi}a|\psi_z(z)|^2$. We keep however the linear dependence of the mean-field interaction energy on the dipole-dipole forces strength $\varepsilon_{dd}$, which originates in the effective interaction potential, $v_{1d}$.

To summarize the above discussion, if a droplet formed  is so large that its density is constant, neglecting boundaries and the surface energy, its energy can be expressed as:
\begin{equation}
 E=\frac{gL_z}{2L^2}\left(1 +\frac{4ac_2}{\pi L} -\varepsilon_{dd}\right)n_{eq}^2+\frac{8\varepsilon_0 c_3 L_z}{\pi^3L}\left(an_{eq}\right)^3,
\label{E_dd}
\end{equation}
where $n_{eq}$ is a saturation density and $L_z$ is the size of the droplet in the $z$ direction. The correction to the mean field coupling (i.e. the s-wave scattering length)  caused by the confinement,  equal to $c_2\xi^2$, is incorporated into the mean field energy.  This way it is easy to notice, that the most important contribution to the LHY energy, stabilizing the system, if the first term becomes negative, is given by  $c_3 \xi^3$. 

The droplet is stable, it neither expands nor  shrinks, \cite{Zin2021}. This is possible if and only if: $(\partial E/\partial L_z)_N=0$  with constraint that the  number of atoms  $N=n_{eq}L_z$ is constant and if the second derivative is positive $(\partial^2 E/\partial^2 L_z)_N>0$. 
Obtained this way equilibrium density is equal to:
 \begin{equation}\label{eqdensity}
 n_{eq}=\frac{L\pi^2}{a^2}\frac{\varepsilon_{dd}-\varepsilon^{cr}_{dd}}{16c_3}. 
 \end{equation}
We notice that  $\varepsilon^{cr}_{dd}=1+4ac_2/(\pi L)$, is the critical value of the dipole-dipole interactions strength, as physically allowed values of  droplet density are  only if $\varepsilon_{dd}> \varepsilon^{cr}_{dd}$. 

Our result resembles the hypothesis of B.~Blakie \cite{bisset2015,blakie2016}, that droplets observed in dysprosium condensates are stabilized by three-body interactions. However, the proposed scenario requires a large elastic three-body cross-section and negligible inelastic processes, thus the hypothesis was soon replaced with the idea of D. Petrov \cite{Petrov2015} pointing to the stabilizing role of quantum fluctuations. In quasi-1D dipolar droplets, three-body interactions indeed stabilize the system; however, they are very small as they affect the energy via quantum fluctuations only.

\section{Conclusions}
\label{sec:conclusions}
The beyond-mean-field correction to the energy of a dipolar gas, confined by an external potential to form a strongly elongated system,  has been discussed. We derived the Lee-Huang-Yang (LHY) energy density term for such a geometry and showed that it can stabilize the gas, preventing it from collapsing in the entire crossover region. It turns out that if dipoles are oriented along the long axis of the system, the anisotropy of the dipolar potential vanishes in the space of reduced dimensionality. The potential becomes very local and attractive. Quasi-1D droplets are formed if the effective 1D dipolar attraction is canceled by repulsion originating from contact interactions. In this region of parameters, the dominant term stabilizing the gas from collapse is the effective three-body interactions present in the quantum fluctuations proportional to the third power of density, $n^3$.  Based on our finding, analysis of numerical data presented in \cite{Edler2017} clearly shows the the similar contribution to the LHY energy is present if the elongated geometry is produced by a harmonic potential. 

We have identified quasi-1D dipolar droplets as promising candidates for studying the influence of three-body fluctuations  on the properties of dilute gases.

\begin{acknowledgments}
PZ and MG  were supported by the Polish National Science Centre through the project MAQS under QuantERA, which has received funding from the European Union’s Horizon 2020 research and innovation program under Grant Agreement No 731473, Project No 2019/32/Z/ST2/00016.
\end{acknowledgments}
\bibliography{refs}

\begin{thebibliography}{50}%
\makeatletter
\providecommand \@ifxundefined [1]{%
 \@ifx{#1\undefined}
}%
\providecommand \@ifnum [1]{%
 \ifnum #1\expandafter \@firstoftwo
 \else \expandafter \@secondoftwo
 \fi
}%
\providecommand \@ifx [1]{%
 \ifx #1\expandafter \@firstoftwo
 \else \expandafter \@secondoftwo
 \fi
}%
\providecommand \natexlab [1]{#1}%
\providecommand \enquote  [1]{``#1''}%
\providecommand \bibnamefont  [1]{#1}%
\providecommand \bibfnamefont [1]{#1}%
\providecommand \citenamefont [1]{#1}%
\providecommand \href@noop [0]{\@secondoftwo}%
\providecommand \href [0]{\begingroup \@sanitize@url \@href}%
\providecommand \@href[1]{\@@startlink{#1}\@@href}%
\providecommand \@@href[1]{\endgroup#1\@@endlink}%
\providecommand \@sanitize@url [0]{\catcode `\\12\catcode `\$12\catcode `\&12\catcode `\#12\catcode `\^12\catcode `\_12\catcode `\%12\relax}%
\providecommand \@@startlink[1]{}%
\providecommand \@@endlink[0]{}%
\providecommand \url  [0]{\begingroup\@sanitize@url \@url }%
\providecommand \@url [1]{\endgroup\@href {#1}{\urlprefix }}%
\providecommand \urlprefix  [0]{URL }%
\providecommand \Eprint [0]{\href }%
\providecommand \doibase [0]{http://dx.doi.org/}%
\providecommand \selectlanguage [0]{\@gobble}%
\providecommand \bibinfo  [0]{\@secondoftwo}%
\providecommand \bibfield  [0]{\@secondoftwo}%
\providecommand \translation [1]{[#1]}%
\providecommand \BibitemOpen [0]{}%
\providecommand \bibitemStop [0]{}%
\providecommand \bibitemNoStop [0]{.\EOS\space}%
\providecommand \EOS [0]{\spacefactor3000\relax}%
\providecommand \BibitemShut  [1]{\csname bibitem#1\endcsname}%
\let\auto@bib@innerbib\@empty
\bibitem [{\citenamefont {Baranov}\ \emph {et~al.}(2012)\citenamefont {Baranov}, \citenamefont {Dalmonte}, \citenamefont {Pupillo},\ and\ \citenamefont {Zoller}}]{baranov2012}%
  \BibitemOpen
  \bibfield  {author} {\bibinfo {author} {\bibfnamefont {M.}~\bibnamefont {Baranov}}, \bibinfo {author} {\bibfnamefont {M.}~\bibnamefont {Dalmonte}}, \bibinfo {author} {\bibfnamefont {G.}~\bibnamefont {Pupillo}}, \ and\ \bibinfo {author} {\bibfnamefont {P.}~\bibnamefont {Zoller}},\ }\href@noop {} {\bibfield  {journal} {\bibinfo  {journal} {Chem. Rev.}\ }\textbf {\bibinfo {volume} {112}},\ \bibinfo {pages} {5012} (\bibinfo {year} {2012})}\BibitemShut {NoStop}%
\bibitem [{\citenamefont {Lahaye}\ \emph {et~al.}(2009)\citenamefont {Lahaye}, \citenamefont {Menotti}, \citenamefont {Santos}, \citenamefont {Lewenstein},\ and\ \citenamefont {Pfau}}]{Lahaye:2009}%
  \BibitemOpen
  \bibfield  {author} {\bibinfo {author} {\bibfnamefont {T.}~\bibnamefont {Lahaye}}, \bibinfo {author} {\bibfnamefont {C.}~\bibnamefont {Menotti}}, \bibinfo {author} {\bibfnamefont {L.}~\bibnamefont {Santos}}, \bibinfo {author} {\bibfnamefont {M.}~\bibnamefont {Lewenstein}}, \ and\ \bibinfo {author} {\bibfnamefont {T.}~\bibnamefont {Pfau}},\ }\href {http://stacks.iop.org/0034-4885/72/i=12/a=126401} {\bibfield  {journal} {\bibinfo  {journal} {Rep. Prog. Phys.}\ }\textbf {\bibinfo {volume} {72}},\ \bibinfo {pages} {126401} (\bibinfo {year} {2009})}\BibitemShut {NoStop}%
\bibitem [{\citenamefont {G\'oral}\ \emph {et~al.}(2000)\citenamefont {G\'oral}, \citenamefont {Rza\ifmmode \mbox{\c{}}\else \c{}\fi{}\ifmmode~\dot{z}\else \.{z}\fi{}ewski},\ and\ \citenamefont {Pfau}}]{Goral2000}%
  \BibitemOpen
  \bibfield  {author} {\bibinfo {author} {\bibfnamefont {K.}~\bibnamefont {G\'oral}}, \bibinfo {author} {\bibfnamefont {K.}~\bibnamefont {Rza\ifmmode \mbox{\c{}}\else \c{}\fi{}\ifmmode~\dot{z}\else \.{z}\fi{}ewski}}, \ and\ \bibinfo {author} {\bibfnamefont {T.}~\bibnamefont {Pfau}},\ }\href {\doibase 10.1103/PhysRevA.61.051601} {\bibfield  {journal} {\bibinfo  {journal} {Phys. Rev. A}\ }\textbf {\bibinfo {volume} {61}},\ \bibinfo {pages} {051601(R)} (\bibinfo {year} {2000})}\BibitemShut {NoStop}%
\bibitem [{\citenamefont {Bohn}\ \emph {et~al.}(2017)\citenamefont {Bohn}, \citenamefont {Rey},\ and\ \citenamefont {Ye}}]{bohn2017cold}%
  \BibitemOpen
  \bibfield  {author} {\bibinfo {author} {\bibfnamefont {J.~L.}\ \bibnamefont {Bohn}}, \bibinfo {author} {\bibfnamefont {A.~M.}\ \bibnamefont {Rey}}, \ and\ \bibinfo {author} {\bibfnamefont {J.}~\bibnamefont {Ye}},\ }\href@noop {} {\bibfield  {journal} {\bibinfo  {journal} {Science}\ }\textbf {\bibinfo {volume} {357}},\ \bibinfo {pages} {1002} (\bibinfo {year} {2017})}\BibitemShut {NoStop}%
\bibitem [{\citenamefont {Saffman}\ \emph {et~al.}(2010)\citenamefont {Saffman}, \citenamefont {Walker},\ and\ \citenamefont {M{\o}lmer}}]{saffman2010quantum}%
  \BibitemOpen
  \bibfield  {author} {\bibinfo {author} {\bibfnamefont {M.}~\bibnamefont {Saffman}}, \bibinfo {author} {\bibfnamefont {T.~G.}\ \bibnamefont {Walker}}, \ and\ \bibinfo {author} {\bibfnamefont {K.}~\bibnamefont {M{\o}lmer}},\ }\href@noop {} {\bibfield  {journal} {\bibinfo  {journal} {Rev. Mod. Phys.}\ }\textbf {\bibinfo {volume} {82}},\ \bibinfo {pages} {2313} (\bibinfo {year} {2010})}\BibitemShut {NoStop}%
\bibitem [{\citenamefont {Lu}\ \emph {et~al.}(2011)\citenamefont {Lu}, \citenamefont {Burdick}, \citenamefont {Youn},\ and\ \citenamefont {Lev}}]{lu2011strongly}%
  \BibitemOpen
  \bibfield  {author} {\bibinfo {author} {\bibfnamefont {M.}~\bibnamefont {Lu}}, \bibinfo {author} {\bibfnamefont {N.~Q.}\ \bibnamefont {Burdick}}, \bibinfo {author} {\bibfnamefont {S.~H.}\ \bibnamefont {Youn}}, \ and\ \bibinfo {author} {\bibfnamefont {B.~L.}\ \bibnamefont {Lev}},\ }\href@noop {} {\bibfield  {journal} {\bibinfo  {journal} {Phys. Rev. Lett.}\ }\textbf {\bibinfo {volume} {107}},\ \bibinfo {pages} {190401} (\bibinfo {year} {2011})}\BibitemShut {NoStop}%
\bibitem [{\citenamefont {Aikawa}\ \emph {et~al.}(2012)\citenamefont {Aikawa}, \citenamefont {Frisch}, \citenamefont {Mark}, \citenamefont {Baier}, \citenamefont {Rietzler}, \citenamefont {Grimm},\ and\ \citenamefont {Ferlaino}}]{aikawa2012bose}%
  \BibitemOpen
  \bibfield  {author} {\bibinfo {author} {\bibfnamefont {K.}~\bibnamefont {Aikawa}}, \bibinfo {author} {\bibfnamefont {A.}~\bibnamefont {Frisch}}, \bibinfo {author} {\bibfnamefont {M.}~\bibnamefont {Mark}}, \bibinfo {author} {\bibfnamefont {S.}~\bibnamefont {Baier}}, \bibinfo {author} {\bibfnamefont {A.}~\bibnamefont {Rietzler}}, \bibinfo {author} {\bibfnamefont {R.}~\bibnamefont {Grimm}}, \ and\ \bibinfo {author} {\bibfnamefont {F.}~\bibnamefont {Ferlaino}},\ }\href@noop {} {\bibfield  {journal} {\bibinfo  {journal} {Phys. Rev. Lett.}\ }\textbf {\bibinfo {volume} {108}},\ \bibinfo {pages} {210401} (\bibinfo {year} {2012})}\BibitemShut {NoStop}%
\bibitem [{\citenamefont {Kadau}\ \emph {et~al.}(2016)\citenamefont {Kadau}, \citenamefont {Schmitt}, \citenamefont {Wenzel}, \citenamefont {Wink}, \citenamefont {Maier}, \citenamefont {Ferrier-Barbut},\ and\ \citenamefont {Pfau}}]{Kadau2016}%
  \BibitemOpen
  \bibfield  {author} {\bibinfo {author} {\bibfnamefont {H.}~\bibnamefont {Kadau}}, \bibinfo {author} {\bibfnamefont {M.}~\bibnamefont {Schmitt}}, \bibinfo {author} {\bibfnamefont {M.}~\bibnamefont {Wenzel}}, \bibinfo {author} {\bibfnamefont {C.}~\bibnamefont {Wink}}, \bibinfo {author} {\bibfnamefont {T.}~\bibnamefont {Maier}}, \bibinfo {author} {\bibfnamefont {I.}~\bibnamefont {Ferrier-Barbut}}, \ and\ \bibinfo {author} {\bibfnamefont {T.}~\bibnamefont {Pfau}},\ }\href@noop {} {\bibfield  {journal} {\bibinfo  {journal} {Nature}\ }\textbf {\bibinfo {volume} {530}},\ \bibinfo {pages} {194} (\bibinfo {year} {2016})}\BibitemShut {NoStop}%
\bibitem [{\citenamefont {Ferrier-Barbut}\ \emph {et~al.}(2016)\citenamefont {Ferrier-Barbut}, \citenamefont {Kadau}, \citenamefont {Schmitt}, \citenamefont {Wenzel},\ and\ \citenamefont {Pfau}}]{Barbut2016}%
  \BibitemOpen
  \bibfield  {author} {\bibinfo {author} {\bibfnamefont {I.}~\bibnamefont {Ferrier-Barbut}}, \bibinfo {author} {\bibfnamefont {H.}~\bibnamefont {Kadau}}, \bibinfo {author} {\bibfnamefont {M.}~\bibnamefont {Schmitt}}, \bibinfo {author} {\bibfnamefont {M.}~\bibnamefont {Wenzel}}, \ and\ \bibinfo {author} {\bibfnamefont {T.}~\bibnamefont {Pfau}},\ }\href {\doibase 10.1103/PhysRevLett.116.215301} {\bibfield  {journal} {\bibinfo  {journal} {Phys. Rev. Lett.}\ }\textbf {\bibinfo {volume} {116}},\ \bibinfo {pages} {215301} (\bibinfo {year} {2016})}\BibitemShut {NoStop}%
\bibitem [{\citenamefont {Chomaz}\ \emph {et~al.}(2016)\citenamefont {Chomaz}, \citenamefont {Baier}, \citenamefont {Petter}, \citenamefont {Mark}, \citenamefont {W{\"a}chtler}, \citenamefont {Santos},\ and\ \citenamefont {Ferlaino}}]{Chomaz2016}%
  \BibitemOpen
  \bibfield  {author} {\bibinfo {author} {\bibfnamefont {L.}~\bibnamefont {Chomaz}}, \bibinfo {author} {\bibfnamefont {S.}~\bibnamefont {Baier}}, \bibinfo {author} {\bibfnamefont {D.}~\bibnamefont {Petter}}, \bibinfo {author} {\bibfnamefont {M.~J.}\ \bibnamefont {Mark}}, \bibinfo {author} {\bibfnamefont {F.}~\bibnamefont {W{\"a}chtler}}, \bibinfo {author} {\bibfnamefont {L.}~\bibnamefont {Santos}}, \ and\ \bibinfo {author} {\bibfnamefont {F.}~\bibnamefont {Ferlaino}},\ }\href@noop {} {\bibfield  {journal} {\bibinfo  {journal} {Phys. Rev. X}\ }\textbf {\bibinfo {volume} {6}},\ \bibinfo {pages} {041039} (\bibinfo {year} {2016})}\BibitemShut {NoStop}%
\bibitem [{\citenamefont {Schmitt}\ \emph {et~al.}(2016)\citenamefont {Schmitt}, \citenamefont {Wenzel}, \citenamefont {B{\"o}ttcher}, \citenamefont {Ferrier-Barbut},\ and\ \citenamefont {Pfau}}]{Schmitt2016}%
  \BibitemOpen
  \bibfield  {author} {\bibinfo {author} {\bibfnamefont {M.}~\bibnamefont {Schmitt}}, \bibinfo {author} {\bibfnamefont {M.}~\bibnamefont {Wenzel}}, \bibinfo {author} {\bibfnamefont {F.}~\bibnamefont {B{\"o}ttcher}}, \bibinfo {author} {\bibfnamefont {I.}~\bibnamefont {Ferrier-Barbut}}, \ and\ \bibinfo {author} {\bibfnamefont {T.}~\bibnamefont {Pfau}},\ }\href@noop {} {\bibfield  {journal} {\bibinfo  {journal} {Nature}\ }\textbf {\bibinfo {volume} {539}},\ \bibinfo {pages} {259} (\bibinfo {year} {2016})}\BibitemShut {NoStop}%
\bibitem [{\citenamefont {Wenzel}\ \emph {et~al.}(2017)\citenamefont {Wenzel}, \citenamefont {B\"ottcher}, \citenamefont {\mbox{Langen}}, \citenamefont {Ferrier-Barbut},\ and\ \citenamefont {Pfau}}]{Wenzel2017}%
  \BibitemOpen
  \bibfield  {author} {\bibinfo {author} {\bibfnamefont {M.}~\bibnamefont {Wenzel}}, \bibinfo {author} {\bibfnamefont {F.}~\bibnamefont {B\"ottcher}}, \bibinfo {author} {\bibfnamefont {T.}~\bibnamefont {\mbox{Langen}}}, \bibinfo {author} {\bibfnamefont {I.}~\bibnamefont {Ferrier-Barbut}}, \ and\ \bibinfo {author} {\bibfnamefont {T.}~\bibnamefont {Pfau}},\ }\href {\doibase 10.1103/PhysRevA.96.053630} {\bibfield  {journal} {\bibinfo  {journal} {Phys. Rev. A}\ }\textbf {\bibinfo {volume} {96}},\ \bibinfo {pages} {053630} (\bibinfo {year} {2017})}\BibitemShut {NoStop}%
\bibitem [{\citenamefont {Petrov}(2015)}]{Petrov2015}%
  \BibitemOpen
  \bibfield  {author} {\bibinfo {author} {\bibfnamefont {D.~S.}\ \bibnamefont {Petrov}},\ }\href@noop {} {\bibfield  {journal} {\bibinfo  {journal} {Phys. Rev. Lett.}\ }\textbf {\bibinfo {volume} {115}},\ \bibinfo {pages} {155302} (\bibinfo {year} {2015})}\BibitemShut {NoStop}%
\bibitem [{\citenamefont {Lee}\ \emph {et~al.}(1957)\citenamefont {Lee}, \citenamefont {Huang},\ and\ \citenamefont {Yang}}]{Lee1957}%
  \BibitemOpen
  \bibfield  {author} {\bibinfo {author} {\bibfnamefont {T.~D.}\ \bibnamefont {Lee}}, \bibinfo {author} {\bibfnamefont {K.}~\bibnamefont {Huang}}, \ and\ \bibinfo {author} {\bibfnamefont {C.~N.}\ \bibnamefont {Yang}},\ }\href@noop {} {\bibfield  {journal} {\bibinfo  {journal} {Physical Review}\ }\textbf {\bibinfo {volume} {106}},\ \bibinfo {pages} {1135} (\bibinfo {year} {1957})}\BibitemShut {NoStop}%
\bibitem [{\citenamefont {Beliaev}(1958)}]{Beliaev1958}%
  \BibitemOpen
  \bibfield  {author} {\bibinfo {author} {\bibfnamefont {S.}~\bibnamefont {Beliaev}},\ }\href@noop {} {\bibfield  {journal} {\bibinfo  {journal} {Sov. Phys. JETP}\ }\textbf {\bibinfo {volume} {34}},\ \bibinfo {pages} {299} (\bibinfo {year} {1958})}\BibitemShut {NoStop}%
\bibitem [{\citenamefont {Schick}(1971)}]{Schick1971}%
  \BibitemOpen
  \bibfield  {author} {\bibinfo {author} {\bibfnamefont {M.}~\bibnamefont {Schick}},\ }\href@noop {} {\bibfield  {journal} {\bibinfo  {journal} {Phys, Rev, A}\ }\textbf {\bibinfo {volume} {3}},\ \bibinfo {pages} {1067} (\bibinfo {year} {1971})}\BibitemShut {NoStop}%
\bibitem [{\citenamefont {Hugenholtz}\ and\ \citenamefont {Pines}(1959)}]{hugenholtz1959ground}%
  \BibitemOpen
  \bibfield  {author} {\bibinfo {author} {\bibfnamefont {N.}~\bibnamefont {Hugenholtz}}\ and\ \bibinfo {author} {\bibfnamefont {D.}~\bibnamefont {Pines}},\ }\href@noop {} {\bibfield  {journal} {\bibinfo  {journal} {Physical Review}\ }\textbf {\bibinfo {volume} {116}},\ \bibinfo {pages} {489} (\bibinfo {year} {1959})}\BibitemShut {NoStop}%
\bibitem [{\citenamefont {Sch\"{u}tzhold}\ \emph {et~al.}(2006)\citenamefont {Sch\"{u}tzhold}, \citenamefont {Uhlmann}, \citenamefont {Xu},\ and\ \citenamefont {\mbox{Fischer}}}]{Uwe2006}%
  \BibitemOpen
  \bibfield  {author} {\bibinfo {author} {\bibfnamefont {R.}~\bibnamefont {Sch\"{u}tzhold}}, \bibinfo {author} {\bibfnamefont {M.}~\bibnamefont {Uhlmann}}, \bibinfo {author} {\bibfnamefont {Y.}~\bibnamefont {Xu}}, \ and\ \bibinfo {author} {\bibfnamefont {U.~R.}\ \bibnamefont {\mbox{Fischer}}},\ }\href {\doibase 10.1142/s0217979206035631} {\bibfield  {journal} {\bibinfo  {journal} {International Journal of Modern Physics B}\ }\textbf {\bibinfo {volume} {20}},\ \bibinfo {pages} {3555} (\bibinfo {year} {2006})}\BibitemShut {NoStop}%
\bibitem [{\citenamefont {Lima}\ and\ \citenamefont {Pelster}(2012)}]{Pelster2012}%
  \BibitemOpen
  \bibfield  {author} {\bibinfo {author} {\bibfnamefont {A.~R.~P.}\ \bibnamefont {Lima}}\ and\ \bibinfo {author} {\bibfnamefont {A.}~\bibnamefont {Pelster}},\ }\href {\doibase 10.1103/PhysRevA.86.063609} {\bibfield  {journal} {\bibinfo  {journal} {Phys. Rev. A}\ }\textbf {\bibinfo {volume} {86}},\ \bibinfo {pages} {063609} (\bibinfo {year} {2012})}\BibitemShut {NoStop}%
\bibitem [{\citenamefont {W{\"a}chtler}\ and\ \citenamefont {Santos}(2016)}]{Wachtler2016}%
  \BibitemOpen
  \bibfield  {author} {\bibinfo {author} {\bibfnamefont {F.}~\bibnamefont {W{\"a}chtler}}\ and\ \bibinfo {author} {\bibfnamefont {L.}~\bibnamefont {Santos}},\ }\href@noop {} {\bibfield  {journal} {\bibinfo  {journal} {Phys. Rev. A}\ }\textbf {\bibinfo {volume} {93}},\ \bibinfo {pages} {061603(R)} (\bibinfo {year} {2016})}\BibitemShut {NoStop}%
\bibitem [{\citenamefont {Bisset}\ \emph {et~al.}(2016)\citenamefont {Bisset}, \citenamefont {Wilson}, \citenamefont {Baillie},\ and\ \citenamefont {Blakie}}]{bisset2016ground}%
  \BibitemOpen
  \bibfield  {author} {\bibinfo {author} {\bibfnamefont {R.~N.}\ \bibnamefont {Bisset}}, \bibinfo {author} {\bibfnamefont {R.~M.}\ \bibnamefont {Wilson}}, \bibinfo {author} {\bibfnamefont {D.}~\bibnamefont {Baillie}}, \ and\ \bibinfo {author} {\bibfnamefont {P.~B.}\ \bibnamefont {Blakie}},\ }\href@noop {} {\bibfield  {journal} {\bibinfo  {journal} {Physical Review A}\ }\textbf {\bibinfo {volume} {94}},\ \bibinfo {pages} {033619} (\bibinfo {year} {2016})}\BibitemShut {NoStop}%
\bibitem [{\citenamefont {Macia}\ \emph {et~al.}(2016)\citenamefont {Macia}, \citenamefont {S{\'a}nchez-Baena}, \citenamefont {Boronat},\ and\ \citenamefont {\mbox{Mazzanti}}}]{Macia2016droplets}%
  \BibitemOpen
  \bibfield  {author} {\bibinfo {author} {\bibfnamefont {A.}~\bibnamefont {Macia}}, \bibinfo {author} {\bibfnamefont {J.}~\bibnamefont {S{\'a}nchez-Baena}}, \bibinfo {author} {\bibfnamefont {J.}~\bibnamefont {Boronat}}, \ and\ \bibinfo {author} {\bibfnamefont {F.}~\bibnamefont {\mbox{Mazzanti}}},\ }\href@noop {} {\bibfield  {journal} {\bibinfo  {journal} {Phys. Rev. Lett.}\ }\textbf {\bibinfo {volume} {117}},\ \bibinfo {pages} {205301} (\bibinfo {year} {2016})}\BibitemShut {NoStop}%
\bibitem [{\citenamefont {Cinti}\ and\ \citenamefont {Boninsegni}(2017)}]{Cinti2017}%
  \BibitemOpen
  \bibfield  {author} {\bibinfo {author} {\bibfnamefont {F.}~\bibnamefont {Cinti}}\ and\ \bibinfo {author} {\bibfnamefont {M.}~\bibnamefont {Boninsegni}},\ }\href@noop {} {\bibfield  {journal} {\bibinfo  {journal} {Phys. Rev. A}\ }\textbf {\bibinfo {volume} {96}},\ \bibinfo {pages} {013627} (\bibinfo {year} {2017})}\BibitemShut {NoStop}%
\bibitem [{\citenamefont {B\"ottcher}\ \emph {et~al.}(2019)\citenamefont {B\"ottcher}, \citenamefont {Wenzel}, \citenamefont {Schmidt}, \citenamefont {Guo}, \citenamefont {\mbox{Langen}}, \citenamefont {Ferrier-Barbut}, \citenamefont {Pfau}, \citenamefont {Bomb\'{\i}n}, \citenamefont {S\'anchez-Baena}, \citenamefont {Boronat},\ and\ \citenamefont {Mazzanti}}]{Bottcher2019a}%
  \BibitemOpen
  \bibfield  {author} {\bibinfo {author} {\bibfnamefont {F.}~\bibnamefont {B\"ottcher}}, \bibinfo {author} {\bibfnamefont {M.}~\bibnamefont {Wenzel}}, \bibinfo {author} {\bibfnamefont {J.-N.}\ \bibnamefont {Schmidt}}, \bibinfo {author} {\bibfnamefont {M.}~\bibnamefont {Guo}}, \bibinfo {author} {\bibfnamefont {T.}~\bibnamefont {\mbox{Langen}}}, \bibinfo {author} {\bibfnamefont {I.}~\bibnamefont {Ferrier-Barbut}}, \bibinfo {author} {\bibfnamefont {T.}~\bibnamefont {Pfau}}, \bibinfo {author} {\bibfnamefont {R.}~\bibnamefont {Bomb\'{\i}n}}, \bibinfo {author} {\bibfnamefont {J.}~\bibnamefont {S\'anchez-Baena}}, \bibinfo {author} {\bibfnamefont {J.}~\bibnamefont {Boronat}}, \ and\ \bibinfo {author} {\bibfnamefont {F.}~\bibnamefont {Mazzanti}},\ }\href {\doibase 10.1103/PhysRevResearch.1.033088} {\bibfield  {journal} {\bibinfo  {journal} {Phys. Rev. Res.}\ }\textbf {\bibinfo {volume} {1}},\ \bibinfo {pages} {033088} (\bibinfo {year} {2019})}\BibitemShut {NoStop}%
\bibitem [{\citenamefont {Sanuy}\ \emph {et~al.}(2024)\citenamefont {Sanuy}, \citenamefont {Bara\ifmmode~\check{c}\else \v{c}\fi{}}, \citenamefont {Stipanovi\ifmmode~\acute{c}\else \'{c}\fi{}}, \citenamefont {Marki\ifmmode~\acute{c}\else \'{c}\fi{}},\ and\ \citenamefont {Boronat}}]{Sanuy2024}%
  \BibitemOpen
  \bibfield  {author} {\bibinfo {author} {\bibfnamefont {A.}~\bibnamefont {Sanuy}}, \bibinfo {author} {\bibfnamefont {R.}~\bibnamefont {Bara\ifmmode~\check{c}\else \v{c}\fi{}}}, \bibinfo {author} {\bibfnamefont {P.}~\bibnamefont {Stipanovi\ifmmode~\acute{c}\else \'{c}\fi{}}}, \bibinfo {author} {\bibfnamefont {L.~V. c.~v.}\ \bibnamefont {Marki\ifmmode~\acute{c}\else \'{c}\fi{}}}, \ and\ \bibinfo {author} {\bibfnamefont {J.}~\bibnamefont {Boronat}},\ }\href {\doibase 10.1103/PhysRevA.109.013313} {\bibfield  {journal} {\bibinfo  {journal} {Phys. Rev. A}\ }\textbf {\bibinfo {volume} {109}},\ \bibinfo {pages} {013313} (\bibinfo {year} {2024})}\BibitemShut {NoStop}%
\bibitem [{\citenamefont {Popov}(1972)}]{Popov}%
  \BibitemOpen
  \bibfield  {author} {\bibinfo {author} {\bibfnamefont {V.~N.}\ \bibnamefont {Popov}},\ }\href@noop {} {\bibfield  {journal} {\bibinfo  {journal} {Theoretical and Mathematical Physics}\ }\textbf {\bibinfo {volume} {11}},\ \bibinfo {pages} {565} (\bibinfo {year} {1972})}\BibitemShut {NoStop}%
\bibitem [{\citenamefont {Mora}\ and\ \citenamefont {Castin}(2009)}]{Mora2009}%
  \BibitemOpen
  \bibfield  {author} {\bibinfo {author} {\bibfnamefont {C.}~\bibnamefont {Mora}}\ and\ \bibinfo {author} {\bibfnamefont {Y.}~\bibnamefont {Castin}},\ }\href@noop {} {\bibfield  {journal} {\bibinfo  {journal} {Phys. Rev. Lett.}\ }\textbf {\bibinfo {volume} {102}},\ \bibinfo {pages} {180404} (\bibinfo {year} {2009})}\BibitemShut {NoStop}%
\bibitem [{\citenamefont {Zin}\ \emph {et~al.}(2021{\natexlab{a}})\citenamefont {Zin}, \citenamefont {Pylak}, \citenamefont {Wasak}, \citenamefont {Jachymski},\ and\ \citenamefont {\mbox{Idziaszek}}}]{Zin_dipole}%
  \BibitemOpen
  \bibfield  {author} {\bibinfo {author} {\bibfnamefont {P.}~\bibnamefont {Zin}}, \bibinfo {author} {\bibfnamefont {M.}~\bibnamefont {Pylak}}, \bibinfo {author} {\bibfnamefont {T.}~\bibnamefont {Wasak}}, \bibinfo {author} {\bibfnamefont {K.}~\bibnamefont {Jachymski}}, \ and\ \bibinfo {author} {\bibfnamefont {Z.}~\bibnamefont {\mbox{Idziaszek}}},\ }\href {\doibase 10.1088/1361-6455/ac2244} {\bibfield  {journal} {\bibinfo  {journal} {Journal of Physics B: Atomic, Molecular and Optical Physics}\ }\textbf {\bibinfo {volume} {54}},\ \bibinfo {pages} {165302} (\bibinfo {year} {2021}{\natexlab{a}})}\BibitemShut {NoStop}%
\bibitem [{\citenamefont {Santos}\ \emph {et~al.}(2003)\citenamefont {Santos}, \citenamefont {Shlyapnikov},\ and\ \citenamefont {Lewenstein}}]{Santos2003}%
  \BibitemOpen
  \bibfield  {author} {\bibinfo {author} {\bibfnamefont {L.}~\bibnamefont {Santos}}, \bibinfo {author} {\bibfnamefont {G.~V.}\ \bibnamefont {Shlyapnikov}}, \ and\ \bibinfo {author} {\bibfnamefont {M.}~\bibnamefont {Lewenstein}},\ }\href@noop {} {\bibfield  {journal} {\bibinfo  {journal} {Phys. Rev. Lett.}\ }\textbf {\bibinfo {volume} {90}},\ \bibinfo {pages} {250403} (\bibinfo {year} {2003})}\BibitemShut {NoStop}%
\bibitem [{\citenamefont {Fischer}(2006)}]{Fischer2006}%
  \BibitemOpen
  \bibfield  {author} {\bibinfo {author} {\bibfnamefont {U.~R.}\ \bibnamefont {Fischer}},\ }\href@noop {} {\bibfield  {journal} {\bibinfo  {journal} {Phys. Rev. A}\ }\textbf {\bibinfo {volume} {73}},\ \bibinfo {pages} {031602(R)} (\bibinfo {year} {2006})}\BibitemShut {NoStop}%
\bibitem [{\citenamefont {Boudjem{\^a}a}\ and\ \citenamefont {Shlyapnikov}(2013)}]{Boudjemaa2013}%
  \BibitemOpen
  \bibfield  {author} {\bibinfo {author} {\bibfnamefont {A.}~\bibnamefont {Boudjem{\^a}a}}\ and\ \bibinfo {author} {\bibfnamefont {G.~V.}\ \bibnamefont {Shlyapnikov}},\ }\href@noop {} {\bibfield  {journal} {\bibinfo  {journal} {Phys. Rev. A}\ }\textbf {\bibinfo {volume} {87}},\ \bibinfo {pages} {025601} (\bibinfo {year} {2013})}\BibitemShut {NoStop}%
\bibitem [{\citenamefont {Chomaz}\ \emph {et~al.}(2018)\citenamefont {Chomaz}, \citenamefont {Bijnen}, \citenamefont {Petter}, \citenamefont {Faraoni}, \citenamefont {Baier}, \citenamefont {Becher}, \citenamefont {Mark}, \citenamefont {Waechtler}, \citenamefont {Santos},\ and\ \citenamefont {Ferlaino}}]{Chomaz2018}%
  \BibitemOpen
  \bibfield  {author} {\bibinfo {author} {\bibfnamefont {L.}~\bibnamefont {Chomaz}}, \bibinfo {author} {\bibfnamefont {R.}~\bibnamefont {Bijnen}}, \bibinfo {author} {\bibfnamefont {D.}~\bibnamefont {Petter}}, \bibinfo {author} {\bibfnamefont {G.}~\bibnamefont {Faraoni}}, \bibinfo {author} {\bibfnamefont {S.}~\bibnamefont {Baier}}, \bibinfo {author} {\bibfnamefont {J.~H.}\ \bibnamefont {Becher}}, \bibinfo {author} {\bibfnamefont {M.~J.}\ \bibnamefont {Mark}}, \bibinfo {author} {\bibfnamefont {F.}~\bibnamefont {Waechtler}}, \bibinfo {author} {\bibfnamefont {L.}~\bibnamefont {Santos}}, \ and\ \bibinfo {author} {\bibfnamefont {F.}~\bibnamefont {Ferlaino}},\ }\href@noop {} {\bibfield  {journal} {\bibinfo  {journal} {Nature Physics}\ }\textbf {\bibinfo {volume} {14}},\ \bibinfo {pages} {442} (\bibinfo {year} {2018})}\BibitemShut {NoStop}%
\bibitem [{\citenamefont {Petter}\ \emph {et~al.}(2019)\citenamefont {Petter}, \citenamefont {Natale}, \citenamefont {van Bijnen}, \citenamefont {\mbox{Patscheider}}, \citenamefont {Mark}, \citenamefont {Chomaz},\ and\ \citenamefont {\mbox{Ferlaino}}}]{Petter2018}%
  \BibitemOpen
  \bibfield  {author} {\bibinfo {author} {\bibfnamefont {D.}~\bibnamefont {Petter}}, \bibinfo {author} {\bibfnamefont {G.}~\bibnamefont {Natale}}, \bibinfo {author} {\bibfnamefont {R.~M.~W.}\ \bibnamefont {van Bijnen}}, \bibinfo {author} {\bibfnamefont {A.}~\bibnamefont {\mbox{Patscheider}}}, \bibinfo {author} {\bibfnamefont {M.~J.}\ \bibnamefont {Mark}}, \bibinfo {author} {\bibfnamefont {L.}~\bibnamefont {Chomaz}}, \ and\ \bibinfo {author} {\bibfnamefont {F.}~\bibnamefont {\mbox{Ferlaino}}},\ }\href {\doibase 10.1103/PhysRevLett.122.183401} {\bibfield  {journal} {\bibinfo  {journal} {Phys. Rev. Lett.}\ }\textbf {\bibinfo {volume} {122}},\ \bibinfo {pages} {183401} (\bibinfo {year} {2019})}\BibitemShut {NoStop}%
\bibitem [{\citenamefont {Kora}\ and\ \citenamefont {Boninsegni}(2019)}]{Kora2019}%
  \BibitemOpen
  \bibfield  {author} {\bibinfo {author} {\bibfnamefont {Y.}~\bibnamefont {Kora}}\ and\ \bibinfo {author} {\bibfnamefont {M.}~\bibnamefont {Boninsegni}},\ }\href@noop {} {\bibfield  {journal} {\bibinfo  {journal} {J. Low. Temp. Phys.}\ }\textbf {\bibinfo {volume} {197}},\ \bibinfo {pages} {337} (\bibinfo {year} {2019})}\BibitemShut {NoStop}%
\bibitem [{\citenamefont {Petrov}\ and\ \citenamefont {Astrakharchik}(2016)}]{Petrov2016}%
  \BibitemOpen
  \bibfield  {author} {\bibinfo {author} {\bibfnamefont {D.~S.}\ \bibnamefont {Petrov}}\ and\ \bibinfo {author} {\bibfnamefont {G.~E.}\ \bibnamefont {Astrakharchik}},\ }\href@noop {} {\bibfield  {journal} {\bibinfo  {journal} {Phys. Rev. Lett.}\ }\textbf {\bibinfo {volume} {117}},\ \bibinfo {pages} {100401} (\bibinfo {year} {2016})}\BibitemShut {NoStop}%
\bibitem [{\citenamefont {Zin}\ \emph {et~al.}(2018)\citenamefont {Zin}, \citenamefont {Pylak}, \citenamefont {Wasak}, \citenamefont {Gajda},\ and\ \citenamefont {Idziaszek}}]{zin2018droplets}%
  \BibitemOpen
  \bibfield  {author} {\bibinfo {author} {\bibfnamefont {P.}~\bibnamefont {Zin}}, \bibinfo {author} {\bibfnamefont {M.}~\bibnamefont {Pylak}}, \bibinfo {author} {\bibfnamefont {T.}~\bibnamefont {Wasak}}, \bibinfo {author} {\bibfnamefont {M.}~\bibnamefont {Gajda}}, \ and\ \bibinfo {author} {\bibfnamefont {Z.}~\bibnamefont {Idziaszek}},\ }\href {\doibase 10.1103/PhysRevA.98.051603} {\bibfield  {journal} {\bibinfo  {journal} {Phys. Rev. A}\ }\textbf {\bibinfo {volume} {98}},\ \bibinfo {pages} {051603(R)} (\bibinfo {year} {2018})}\BibitemShut {NoStop}%
\bibitem [{\citenamefont {Ilg}\ \emph {et~al.}(2018)\citenamefont {Ilg}, \citenamefont {Kumlin}, \citenamefont {Santos}, \citenamefont {Petrov},\ and\ \citenamefont {B\"uchler}}]{buechler2018crossover}%
  \BibitemOpen
  \bibfield  {author} {\bibinfo {author} {\bibfnamefont {T.}~\bibnamefont {Ilg}}, \bibinfo {author} {\bibfnamefont {J.}~\bibnamefont {Kumlin}}, \bibinfo {author} {\bibfnamefont {L.}~\bibnamefont {Santos}}, \bibinfo {author} {\bibfnamefont {D.~S.}\ \bibnamefont {Petrov}}, \ and\ \bibinfo {author} {\bibfnamefont {H.~P.}\ \bibnamefont {B\"uchler}},\ }\href {\doibase 10.1103/PhysRevA.98.051604} {\bibfield  {journal} {\bibinfo  {journal} {Phys. Rev. A}\ }\textbf {\bibinfo {volume} {98}},\ \bibinfo {pages} {051604(R)} (\bibinfo {year} {2018})}\BibitemShut {NoStop}%
\bibitem [{\citenamefont {Zin}\ \emph {et~al.}(2022{\natexlab{a}})\citenamefont {Zin}, \citenamefont {Pylak},\ and\ \citenamefont {Gajda}}]{Zin2022a}%
  \BibitemOpen
  \bibfield  {author} {\bibinfo {author} {\bibfnamefont {P.}~\bibnamefont {Zin}}, \bibinfo {author} {\bibfnamefont {M.}~\bibnamefont {Pylak}}, \ and\ \bibinfo {author} {\bibfnamefont {M.}~\bibnamefont {Gajda}},\ }\href {\doibase 10.1103/PhysRevA.106.013320} {\bibfield  {journal} {\bibinfo  {journal} {Phys. Rev. A}\ }\textbf {\bibinfo {volume} {106}},\ \bibinfo {pages} {013320} (\bibinfo {year} {2022}{\natexlab{a}})}\BibitemShut {NoStop}%
\bibitem [{\citenamefont {Edler}\ \emph {et~al.}(2017)\citenamefont {Edler}, \citenamefont {Mishra}, \citenamefont {W\"achtler}, \citenamefont {Nath}, \citenamefont {Sinha},\ and\ \citenamefont {Santos}}]{Edler2017}%
  \BibitemOpen
  \bibfield  {author} {\bibinfo {author} {\bibfnamefont {D.}~\bibnamefont {Edler}}, \bibinfo {author} {\bibfnamefont {C.}~\bibnamefont {Mishra}}, \bibinfo {author} {\bibfnamefont {F.}~\bibnamefont {W\"achtler}}, \bibinfo {author} {\bibfnamefont {R.}~\bibnamefont {Nath}}, \bibinfo {author} {\bibfnamefont {S.}~\bibnamefont {Sinha}}, \ and\ \bibinfo {author} {\bibfnamefont {L.}~\bibnamefont {Santos}},\ }\href {\doibase 10.1103/PhysRevLett.119.050403} {\bibfield  {journal} {\bibinfo  {journal} {Phys. Rev. Lett.}\ }\textbf {\bibinfo {volume} {119}},\ \bibinfo {pages} {050403} (\bibinfo {year} {2017})}\BibitemShut {NoStop}%
\bibitem [{\citenamefont {Zin}\ \emph {et~al.}(2022{\natexlab{b}})\citenamefont {Zin}, \citenamefont {Pylak}, \citenamefont {Idziaszek},\ and\ \citenamefont {Gajda}}]{Zin2022b}%
  \BibitemOpen
  \bibfield  {author} {\bibinfo {author} {\bibfnamefont {P.}~\bibnamefont {Zin}}, \bibinfo {author} {\bibfnamefont {M.}~\bibnamefont {Pylak}}, \bibinfo {author} {\bibfnamefont {Z.}~\bibnamefont {Idziaszek}}, \ and\ \bibinfo {author} {\bibfnamefont {M.}~\bibnamefont {Gajda}},\ }\href {\doibase 10.1088/1367-2630/aca175} {\bibfield  {journal} {\bibinfo  {journal} {New Journal of Physics}\ }\textbf {\bibinfo {volume} {24}},\ \bibinfo {pages} {113038} (\bibinfo {year} {2022}{\natexlab{b}})}\BibitemShut {NoStop}%
\bibitem [{\citenamefont {Yi}\ and\ \citenamefont {You}(2000)}]{Yi2000}%
  \BibitemOpen
  \bibfield  {author} {\bibinfo {author} {\bibfnamefont {S.}~\bibnamefont {Yi}}\ and\ \bibinfo {author} {\bibfnamefont {L.}~\bibnamefont {You}},\ }\href {\doibase 10.1103/PhysRevA.61.041604} {\bibfield  {journal} {\bibinfo  {journal} {Phys. Rev. A}\ }\textbf {\bibinfo {volume} {61}},\ \bibinfo {pages} {041604(R)} (\bibinfo {year} {2000})}\BibitemShut {NoStop}%
\bibitem [{\citenamefont {O\l{}dziejewski}\ and\ \citenamefont {Jachymski}(2016)}]{Oldziejewski2016}%
  \BibitemOpen
  \bibfield  {author} {\bibinfo {author} {\bibfnamefont {R.}~\bibnamefont {O\l{}dziejewski}}\ and\ \bibinfo {author} {\bibfnamefont {K.}~\bibnamefont {Jachymski}},\ }\href {\doibase 10.1103/PhysRevA.94.063638} {\bibfield  {journal} {\bibinfo  {journal} {Phys. Rev. A}\ }\textbf {\bibinfo {volume} {94}},\ \bibinfo {pages} {063638} (\bibinfo {year} {2016})}\BibitemShut {NoStop}%
\bibitem [{\citenamefont {Abad}\ \emph {et~al.}(2009)\citenamefont {Abad}, \citenamefont {Guilleumas}, \citenamefont {Mayol}, \citenamefont {Pi},\ and\ \citenamefont {Jezek}}]{Abad2020}%
  \BibitemOpen
  \bibfield  {author} {\bibinfo {author} {\bibfnamefont {M.}~\bibnamefont {Abad}}, \bibinfo {author} {\bibfnamefont {M.}~\bibnamefont {Guilleumas}}, \bibinfo {author} {\bibfnamefont {R.}~\bibnamefont {Mayol}}, \bibinfo {author} {\bibfnamefont {M.}~\bibnamefont {Pi}}, \ and\ \bibinfo {author} {\bibfnamefont {D.~M.}\ \bibnamefont {Jezek}},\ }\href {\doibase 10.1103/PhysRevA.79.063622} {\bibfield  {journal} {\bibinfo  {journal} {Phys. Rev. A}\ }\textbf {\bibinfo {volume} {79}},\ \bibinfo {pages} {063622} (\bibinfo {year} {2009})}\BibitemShut {NoStop}%
\bibitem [{\citenamefont {Ota}\ and\ \citenamefont {Astrakharchik}(2020)}]{Ota2020}%
  \BibitemOpen
  \bibfield  {author} {\bibinfo {author} {\bibfnamefont {M.}~\bibnamefont {Ota}}\ and\ \bibinfo {author} {\bibfnamefont {G.~E.}\ \bibnamefont {Astrakharchik}},\ }\href {\doibase 10.21468/SciPostPhys.9.2.020} {\bibfield  {journal} {\bibinfo  {journal} {SciPost Phys.}\ }\textbf {\bibinfo {volume} {9}},\ \bibinfo {pages} {20} (\bibinfo {year} {2020})}\BibitemShut {NoStop}%
\bibitem [{\citenamefont {Hu}\ and\ \citenamefont {Liu}(2020)}]{Hu2020}%
  \BibitemOpen
  \bibfield  {author} {\bibinfo {author} {\bibfnamefont {H.}~\bibnamefont {Hu}}\ and\ \bibinfo {author} {\bibfnamefont {X.-J.}\ \bibnamefont {Liu}},\ }\href {\doibase 10.1103/PhysRevLett.125.195302} {\bibfield  {journal} {\bibinfo  {journal} {Phys. Rev. Lett.}\ }\textbf {\bibinfo {volume} {125}},\ \bibinfo {pages} {195302} (\bibinfo {year} {2020})}\BibitemShut {NoStop}%
\bibitem [{\citenamefont {Olshanii}(1998)}]{Olshanii98}%
  \BibitemOpen
  \bibfield  {author} {\bibinfo {author} {\bibfnamefont {M.}~\bibnamefont {Olshanii}},\ }\href {\doibase 10.1103/PhysRevLett.81.938} {\bibfield  {journal} {\bibinfo  {journal} {Phys. Rev. Lett.}\ }\textbf {\bibinfo {volume} {81}},\ \bibinfo {pages} {938} (\bibinfo {year} {1998})}\BibitemShut {NoStop}%
\bibitem [{\citenamefont {Bergeman}\ \emph {et~al.}(2003)\citenamefont {Bergeman}, \citenamefont {Moore},\ and\ \citenamefont {Olshanii}}]{Olshanii03}%
  \BibitemOpen
  \bibfield  {author} {\bibinfo {author} {\bibfnamefont {T.}~\bibnamefont {Bergeman}}, \bibinfo {author} {\bibfnamefont {M.~G.}\ \bibnamefont {Moore}}, \ and\ \bibinfo {author} {\bibfnamefont {M.}~\bibnamefont {Olshanii}},\ }\href {\doibase 10.1103/PhysRevLett.91.163201} {\bibfield  {journal} {\bibinfo  {journal} {Phys. Rev. Lett.}\ }\textbf {\bibinfo {volume} {91}},\ \bibinfo {pages} {163201} (\bibinfo {year} {2003})}\BibitemShut {NoStop}%
\bibitem [{\citenamefont {Zin}\ \emph {et~al.}(2021{\natexlab{b}})\citenamefont {Zin}, \citenamefont {Pylak},\ and\ \citenamefont {Gajda}}]{Zin2021}%
  \BibitemOpen
  \bibfield  {author} {\bibinfo {author} {\bibfnamefont {P.}~\bibnamefont {Zin}}, \bibinfo {author} {\bibfnamefont {M.}~\bibnamefont {Pylak}}, \ and\ \bibinfo {author} {\bibfnamefont {M.}~\bibnamefont {Gajda}},\ }\href {\doibase 10.1103/PhysRevA.103.013312} {\bibfield  {journal} {\bibinfo  {journal} {Phys. Rev. A}\ }\textbf {\bibinfo {volume} {103}},\ \bibinfo {pages} {013312} (\bibinfo {year} {2021}{\natexlab{b}})}\BibitemShut {NoStop}%
\bibitem [{\citenamefont {Bisset}\ and\ \citenamefont {Blakie}(2015)}]{bisset2015}%
  \BibitemOpen
  \bibfield  {author} {\bibinfo {author} {\bibfnamefont {R.~N.}\ \bibnamefont {Bisset}}\ and\ \bibinfo {author} {\bibfnamefont {P.~B.}\ \bibnamefont {Blakie}},\ }\href {\doibase 10.1103/PhysRevA.92.061603} {\bibfield  {journal} {\bibinfo  {journal} {Phys. Rev. A}\ }\textbf {\bibinfo {volume} {92}},\ \bibinfo {pages} {061603(R)} (\bibinfo {year} {2015})}\BibitemShut {NoStop}%
\bibitem [{\citenamefont {Blakie}(2016)}]{blakie2016}%
  \BibitemOpen
  \bibfield  {author} {\bibinfo {author} {\bibfnamefont {P.~B.}\ \bibnamefont {Blakie}},\ }\href {\doibase 10.1103/PhysRevA.93.033644} {\bibfield  {journal} {\bibinfo  {journal} {Phys. Rev. A}\ }\textbf {\bibinfo {volume} {93}},\ \bibinfo {pages} {033644} (\bibinfo {year} {2016})}\BibitemShut {NoStop}%
\end{thebibliography}%
\bibliographystyle{apsrev4-1}

\appendix

\section{Evaluation of the Fourier transform of the dipole-dipole interaction potential}
\label{app0}

Potential of a dipole in the   3D space takes the form: $v_{dip}(\x) = g \varepsilon_{dd} \frac{3 }{4\pi} \frac{1 - 3 ({\bf e}_\x \cdot {\bf e})^2}{r^3}$ (where ${\bf e}_\x = \x/|\x|$ and ${\bf e}$ is the direction of dipole moment). To evaluate its Fourier transform, because the potential is singular at $r=0$,  some regularization procedure is needed.  It can be done by taking the integral in a finite space region between two spheres of radius $r_0$ and $R_0$. The limits $r_0 \rightarrow 0$ and $R_0 \rightarrow \infty$ give the  Fourier decomposition,  $ v^{3d}_{dip}(\K) =  \varepsilon_{dd} \left( 3 (\K \cdot {\bf e})^2/k^2-1 \right) $.  In our case, we need to obtain the dipolar potential in a space constrained to an $L \times L$ box in the $x$-$y$ plane. We define this potential via the Fourier transform of the 3D dipolar potential, multiplied by a mask—specifically, the characteristic function of the region where the atoms are confined: 
\begin{equation}\label{ap11}
  v_{dip}(\K) = \int_{L \times L} \mbox{d} \x_\perp  \int \mbox{d} z \, e^{-i\K\x} \frac{v_{dip}(\x)}{g},
\end{equation}
where $k_{x,y} = \frac{2\pi}{L} q_{x,y}$ are quantized. The upper limit of the integral is explicit, however  we should take into account the  cut-off at $r_0$,  to regularize the integral at small distances, $\x=0$.
This can be done by introducing a mask $h(k)$ in the momentum space:
\begin{equation}\label{ap12}
  v_{dip}(\x) = \frac{1}{(2\pi)^3}\int \mbox{d} \K' \, e^{i \K' \x} v^{3d}_{dip}(\K') h(k') .
\end{equation}
where $h(k')=1$ for $k'<k_0$ and goes to zero otherwise. Parameter $k_0 \simeq 1/r_0$, and at the end we take the limit $k_0 \rightarrow \infty$. Inserting Eq.~(\ref{ap12}) into Eq.~(\ref{ap11}) leads to:
\begin{eqnarray}
  v_{dip}(\K) &=& \frac{1}{(2\pi)^3} \int_{L\times L} \mbox{d}  \x_\perp \int \mbox{d} z  \, 
  e^{-i\K\x}  \nonumber \\ 
 && \int \mbox{d} \K' \, e^{i \K'\x} v^{3d}_{dip}(\K') h(k').
\end{eqnarray}
Analytical integration gives:
\begin{eqnarray}
  v_{dip}(\K) & = & \frac{1}{(2\pi)^2} \int_{L \times L} \mbox{d} \x_\perp \int \mbox{d} \K_\perp' \, e^{-i(\K_\perp-\K_\perp')\x_\perp} \nonumber \\
  &\times& v^{3d}_{dip}(\K_\perp',k_z) 
  h \left(\sqrt{{\K_\perp'}^2+k_z^2} \right).
\end{eqnarray}
In the considered geometry, i.e. if dipoles are oriented along the $z$ axis, we have  $v^{3d}_{dip}(\K) = \varepsilon_{dd}(3k_z^2/k^2-1)$, so we have:
\begin{eqnarray}
v_{dip}(\K) &=& \frac{ \varepsilon_{dd}}{(2\pi)^2} \int_{L \times L}\mbox{d} \x_\perp \int \mbox{d} \K_\perp' \, e^{-i(\K_\perp-\K_\perp')\x_\perp} \nonumber \\
&\times& \left(  \frac{3  k_z^2}{{k_\perp'}^2 + {k_z}^2} -1 \right)
  h (\sqrt{{k_\perp'}^2+{k_z}^2} ).
 \end{eqnarray}
Two types of integrals appears above. The value of the first one, $I_1$ goes to one:
\begin{eqnarray}
I_1&=&  \frac{1}{(2\pi)^2} \int_{L \times L}\hspace{-1em}\mbox{d} \x_\perp \int \mbox{d} \K_\perp' \, e^{-i(\K_\perp-\K_\perp')\x_\perp}  \nonumber \\
&\times& h \left(\sqrt{{k_\perp'}^2+{k_z}^2} \right)\rightarrow 1.
\end{eqnarray}
 The above can be noticed by choosing  $h(k') = \exp(-{k'}^2/k_0^2)$, performing the integrals analytically, and finally taking the desired limit. The second integral, $I_2$ takes the form :
\begin{eqnarray}
I_2&=&\frac{1}{(2\pi)^3}\int_{-L/2}^{L/2}{d\x_\perp}\int d\K_\perp'\, e^{-i(\K_\perp-\K'_\perp)\x_\perp} \\
&\times& \frac{3k_z^2}{k_x'^2+k_y'^2k_z^2}h\left(\sqrt{k_x'^2+k_y'^2+k_z^2}\right),
\end{eqnarray}
Integration with respect to variables  $x$ and $y$ can be performed analytically. After changing variables: $k_\alpha=\frac{2\pi}{L}q_\alpha$ and $k'_\alpha=\frac{2\pi}{L}q'_\alpha$ where $\alpha=\{x,y,z\}$ and $q_{x,y}$ are integers, we obtain:
\begin{widetext}
\begin{equation}
I_2=\frac{3q_z^2}{\pi^2}\int dq_x'\,dq_y'\, \frac{\sin{\pi(q_x-q_x')}}{q_x-q_x'}\frac{\sin{\pi(q_y-q_y')}}{q_y-q_y'}  \frac{1}{q_x'^2+q_y'^2+q_z^2}h\left(\sqrt{q_x'^2+q_y'^2+q_z^2}\right).
\end{equation}
Next, integrating over  $q_y'$ gives the result:
\begin{equation}
  I_2 =  \frac{3q_z^2}{\pi}\int dq_x'\, \frac{\sin{\pi(q_x-q_x')}}{q_x-q_x'}  
  \frac{1-(-1)^{q_y}\exp{\left(-\pi\sqrt{q_x'^2+q_z^2}\right)}}{q_x'^2+q_y^2+q_z^2} h\left(\sqrt{q_x'^2+q_z^2}\right).
\end{equation}
\end{widetext}
The last integral  cannot be evaluated analytically so numerical calculation is necessary. 


\section{Evaluation of the LHY term in the quasi-1D limit}
\label{app:eLHY1d}
The Lee-Huang-Yang energy density, given by Eq. (\ref{eq:lhydef}) can be expressed as follows:
\begin{equation}
    e_{LHY}^{1d}=-\frac{\xi^2}{2}\left(g(\xi)-2c_2\right),
\end{equation}
where:
\begin{eqnarray}
    g(\xi)&=&\hspace{-0.5em}\sum_{q_x,q_y}\int{dq_z\left( \frac{v^2(\q)}{ \varepsilon(\q) + q^2 + 
  \xi v(\q)}-\frac{v^2(\q)}{2q^2}\right)},\\
  c_2&=&-\frac{1}{2}\left(\sum_{q_x,q_y}\int{dq_z\,\frac{v^2(\q)}{2q^2}}-\int{d\q\,\frac{v_{3d}^2(\q)}{2q^2}}\right).
\end{eqnarray}
$c_2$ is a  constant: $c_2\simeq-0.479$ and $g(\xi)$ for $\xi\ll1$ can be expanded into series:
\begin{equation}
    g(\xi)\approx-\xi\sum_{q_x,q_y}\int{dq_z\,\frac{v^3(\q)}{2q^4}}=-2\xi c_3.
\end{equation}
The above integral is well defined. It turns out the for $q_x=q_y=0$, and at the critical point $\varepsilon_{dd}=1$, the $q_z$ dependence of the potential $v(\q)$ is stronger than linear and weaker than quadratic. This makes the LHY energy also to be well defined for all values of the parameter $\xi$ (proportional to the density). 

The coefficient $c_3$ must be calculated in two steps. First, for $|q_z| \leq q_c$, the numerical integration present in the definition of $v(\q)$ needs to be performed. The series over $q_x$ and $q_y$ converges quickly, so only a few terms need to be included. Second, for $|q_z| > q_c$, the interaction potential $v(\q)$ can be approximated as:
\begin{widetext}
\begin{eqnarray}
 v(\q) & = & \frac{3q_z^2}{\pi}\int dq_x'\, \frac{\sin{(\pi(q_x-q_x')}}{q_x-q_x'} 
 \frac{1-(-1)^{q_y}\exp{ \left(-\pi\sqrt{q_x'^2+q_z^2}\right) } }{q_x'^2+q_y^2+q_z^2} 
 = \frac{3q_z^2}{q^2}
 \left(1-(-1)^{q_x}e^{-\pi\sqrt{q_y^2+q_z^2}}\right),
\end{eqnarray}    
\end{widetext}
which allows for analytical integration and summation present in the definition of $c_3$:
\begin{eqnarray}
    2c_3=\sum_{q_x,q_y}\int_{-q_c}^{q_c}{dq_z\,\frac{v^3(\q)}{2q^4}}+\frac{27\pi}{4q_c}.
\end{eqnarray}
Finally, we get $c_3\simeq 84.5301$.
The momentum $q_c$ separating regions of small and large wavectors has to be sufficiently large, i.e. $q_c > 1$. It means that corresponding wavelength is smaller then the confinement extension. Let us observe that for $q_z< q_c$  we get exact result up to numerical accuracy. The second term is given by the approximate but analytic formula, and this term vanishes as $1/q_c$ with increasing $q_c$. Sum of these two contribution depends only very weakly on $q_c$ and saturates for large $q_c$. We checked numerically, that for $q_c >5$, the result is practically constant.

\section{Interaction energy}
\label{App:D}

The interaction energy of the  system confined  within the square prism interacting by the {\it periodic}  potential, $v(\x)$ is given by: 
\begin{eqnarray} 
\label{A-en}
E_{int}^{per} & = & \frac{1}{2}\int  dz \int dz'  \int_{L\times L}  d\x_\perp  \int_{L\times L}  d\x_\perp'   \nonumber \\
&\times&  v(\x-\x') |\psi(\x)|^2 |\psi(\x')|^2,
  \label{eq:energiaint}
\end{eqnarray}
where the periodic potential is given by:
\begin{eqnarray}
\label{D-1}
v(\x) = \frac{g}{2 \pi L^2} \sum_{k_x,k_y} e^{ik_x x + i k_y y} \int d k_z \, e^{i k_z z} v(\K),
\end{eqnarray}
where $v(\K)$ is given by  Eq.~(\ref{eq:vq}).

Now we consider the case of uniform density in the $x$-$y$ plane and use the potential $v_{1d}$ defined by Eq.~(\ref{v1d:def}):
\begin{equation}
    v_{1d}(z-z')=\frac{1}{L^4}\int_{L\times L}{v(\x-\x')\,d\x_\perp\,d\x_\perp'},
\end{equation}
to get: 
\begin{equation} 
E_{int}^{per}  =  \frac{1}{2}\int  dz \int dz' \,  v_{1d}(z-z') |\psi_z(z)|^2 |\psi_z(z')|^2.
\end{equation}
Now we define $n_z(k_z)$ through equation
\begin{eqnarray}
\label{D-3}
n_z(z) = \frac{1}{2 \pi}  \int d k_z \, e^{i k_z z} n_z(k_z).
\end{eqnarray} 
From the above we have:
\begin{eqnarray}  \label{Eperint}
E^{per}_{int} =
\frac{1}{2}
\frac{g}{2 \pi L^2} \int d k_z \, v(k_z) |n(k_z)|^2
\end{eqnarray}

We now make use of Eq.~(\ref{eq:vq}) rewriting it as
\begin{equation}\label{vkz}
v(k_z) = 1 - \varepsilon_{dd}+  \delta v(k_z)
\end{equation}
where $\delta v(k_z)$ goes to zero as $k_z$ goes to zero.
Inserting Eq.~(\ref{vkz}) into Eq.~(\ref{Eperint}) we obtain
\begin{equation} \label{Eper}
E^{per}_{int} = \frac{1}{2}
\frac{g}{2 \pi L^2}  \int d k_z \, \left( 1 - \varepsilon_{dd}+    \delta v(k_z) \right) |n(k_z)|^2 
\end{equation}
If the droplet size in the $z$ direction grows, the integral $\int d k_z \, \delta v(k_z) |n(k_z)|^2  $ decreases. 
Thus for large droplets we may approximate
\begin{equation}
    v_{1d}(z-z')=\frac{g}{L^2}\left( 1 - \varepsilon_{dd} \right)\delta(z-z').
\end{equation}
The above result also shows that the uniform solution is stable for $\varepsilon_{dd} \leq 1$.

We consider now, the energy of interaction if the two-body potential is given by $v_{3d}(\x)$:  \begin{eqnarray} 
\label{D-4}
E^{3d}_{int} &=& \frac{1}{2}\int  dz \int dz' \int_{L\times L} d\x_\perp  \int_{L\times L}  d\x_\perp' \nonumber \\
&\times& v_{3d}(\x-\x') |\psi(\x)|^2 |\psi(\x')|^2,
\end{eqnarray}

Substituting  Eqs.(\ref{D-3}) and (\ref{eq:fourier}) into Eq.~(\ref{D-4}) we get:
\begin{eqnarray}
\label{C11}
    E^{3d}_{int}&=&\frac{g}{2(2\pi)^3} \int d\mbox{\K}  \left(\frac{\sin(k_x L/2)}{k_x L/2} \right)^2 \left(\frac{\sin(k_y L/2)}{k_y L/2} \right)^2 \nonumber\\
&&\left( (1 - \varepsilon_{dd}) + 3\varepsilon_{dd} \frac{k_z^2}{k_x^2+k_y^2+k_z^2}  \right) |n(k_z)|^2. 
\end{eqnarray}

We find that the contribution to the energy originating in the first term in Eq~(\ref{C11}) is proportional to:
\begin{eqnarray*}
\frac{1 - \varepsilon_{dd}}{(2\pi)^2} \int d \K_\perp \,  \left(\frac{\sin(k_x L/2)}{k_x L/2} \right)^2 \left(\frac{\sin(k_y L/2)}{k_y L/2} \right)^2
=\frac{1 - \varepsilon_{dd}}{L^2},
\end{eqnarray*}
while the contribution from the second term, denoted by $\delta v_{3d}(kz)$:
\begin{widetext}
\begin{equation}
    \delta v_{3d}(kz) =  \frac{1}{(2\pi)^2} \int d \K_\perp \, \left(\frac{\sin(k_x L/2)}{k_x L/2} \right)^2 \left(\frac{\sin(k_y L/2)}{k_y L/2} \right)^2 3\varepsilon_{dd} \frac{k_z^2}{k_x^2+k_y^2+k_z^2}, 
\end{equation}
\end{widetext}
goes to zero $\delta v_{3d}(k_z) \to 0$, if  $k_z \to 0$. 
Using the above we have
\begin{equation}\label{eqE3d}
    E^{3d}_{int}=\frac{g}{2(2\pi)} \int d k_z \,  
\left( (1 - \varepsilon_{dd}) +\delta v_{3d}(k_z) \right) |n(k_z)|^2 .
\end{equation}
By comparing  Eqs.~(\ref{Eper}) and (\ref{eqE3d}) we find that both  these equations give the same mean-field energy of interaction for wide droplets. This is the case if  both $\delta v(k_z)$ and $\delta v_{3d}(k_z)$ go to zero as $k_z$ goes to zero.

\end{document}